\documentclass[twocolumn,showpacs,superscriptaddress,floatfix,longbibliography,PRB]{revtex4}
\usepackage{graphicx,amsfonts,amssymb,amsmath,xspace}
\usepackage[colorlinks=true,citecolor=green,linkcolor=red,urlcolor=blue]{hyperref}
\usepackage{bbm}
\usepackage{empheq}

\usepackage{color}

\newcommand{\be}{\begin{equation}}
\newcommand{\ee}{\end{equation}}         
\newcommand{\etal}{{\it et al.}}

\def\nbC{{\mathchoice {\setbox0=\hbox{$\displaystyle\rm C$}%
\hbox{\hbox to0pt{\kern0.4\wd0\vrule height0.9\ht0\hss}\box0}} 
{\setbox0=\hbox{$\textstyle\rm
C$}\hbox{\hbox to0pt{\kern0.4\wd0\vrule height0.9\ht0\hss}\box0}} 
{\setbox0=\hbox{$\scriptstyle\rm
C$}\hbox{\hbox to0pt{\kern0.4\wd0\vrule height0.9\ht0\hss}\box0}}
{\setbox0=\hbox{$\scriptscriptstyle\rm C$}\hbox{\hbox to0pt{\kern0.4\wd0\vrule
height0.9\ht0\hss}\box0}}}}
\def\nbQ{{\mathchoice {\setbox0=\hbox{$\displaystyle\rm 
Q$}\hbox{\raise 0.15\ht0\hbox
to0pt{\kern0.4\wd0\vrule height0.8\ht0\hss}\box0}} 
{\setbox0=\hbox{$\textstyle\rm Q$}\hbox{\raise
0.15\ht0\hbox to0pt{\kern0.4\wd0\vrule height0.8\ht0\hss}\box0}} 
{\setbox0=\hbox{$\scriptstyle\rm
Q$}\hbox{\raise 0.15\ht0\hbox to0pt{\kern0.4\wd0\vrule 
height0.7\ht0\hss}\box0}}
{\setbox0=\hbox{$\scriptscriptstyle\rm Q$}\hbox{\raise 0.15\ht0\hbox 
to0pt{\kern0.4\wd0\vrule
height0.7\ht0\hss}\box0}}}}
\def\nbT{{\mathchoice {\setbox0=\hbox{$\displaystyle\rm 
T$}\hbox{\hbox to0pt{\kern0.3\wd0\vrule
height0.9\ht0\hss}\box0}} {\setbox0=\hbox{$\textstyle\rm 
T$}\hbox{\hbox to0pt{\kern0.3\wd0\vrule
height0.9\ht0\hss}\box0}} {\setbox0=\hbox{$\scriptstyle\rm 
T$}\hbox{\hbox to0pt{\kern0.3\wd0\vrule
height0.9\ht0\hss}\box0}} {\setbox0=\hbox{$\scriptscriptstyle\rm T$}\hbox{\hbox
to0pt{\kern0.3\wd0\vrule height0.9\ht0\hss}\box0}}}}
\def\nbS{{\mathchoice {\setbox0=\hbox{$\displaystyle     \rm 
S$}\hbox{\raise0.5\ht0%
\hbox to0pt{\kern0.35\wd0\vrule height0.45\ht0\hss}\hbox 
to0pt{\kern0.55\wd0\vrule
height0.5\ht0\hss}\box0}} {\setbox0=\hbox{$\textstyle        \rm 
S$}\hbox{\raise0.5\ht0%
\hbox to0pt{\kern0.35\wd0\vrule height0.45\ht0\hss}\hbox 
to0pt{\kern0.55\wd0\vrule
height0.5\ht0\hss}\box0}} {\setbox0=\hbox{$\scriptstyle      \rm 
S$}\hbox{\raise0.5\ht0%
\hboxto0pt{\kern0.35\wd0\vrule height0.45\ht0\hss}\raise0.05\ht0%
\hbox to0pt{\kern0.5\wd0\vrule height0.45\ht0\hss}\box0}} 
{\setbox0=\hbox{$\scriptscriptstyle\rm
S$}\hbox{\raise0.5\ht0%
\hboxto0pt{\kern0.4\wd0\vrule height0.45\ht0\hss}\raise0.05\ht0%
\hbox to0pt{\kern0.55\wd0\vrule height0.45\ht0\hss}\box0}}}}
\def\nbZ{{\mathchoice {\hbox{$\sf\textstyle Z\kern-0.4em Z$}} 
{\hbox{$\sf\textstyle Z\kern-0.4em Z$}}
{\hbox{$\sf\scriptstyle Z\kern-0.3em Z$}} 
{\hbox{$\sf\scriptscriptstyle Z\kern-0.2em Z$}}}}

\begin{document}

\title{{The flat phase of  quantum polymerized membranes}}

\author{O. Coquand} \email{coquand@lptmc.jussieu.fr}
\affiliation{Sorbonne Universit\'es, UPMC Univ  Paris 06,  LPTMC,  CNRS  UMR 7600, F-75005, Paris, France}

\author{D. Mouhanna} \email{mouhanna@lptmc.jussieu.fr}
\affiliation{Sorbonne Universit\'es, UPMC Univ  Paris 06,  LPTMC,  CNRS  UMR 7600, F-75005, Paris, France}

\begin{abstract}

We  investigate   the flat  phase  of  quantum polymerized  phantom membranes by means of a nonperturbative renormalization group  approach.   We first implement this   formalism for  general quantum polymerized  membranes and  derive the flow  equations that   encompass  both quantum and   thermal   fluctuations.  We  then deduce and analyze   the  flow   equations   relevant to study the flat phase  and discuss  their  salient features : quantum to classical crossover and, in each of these regimes,  strong to weak coupling  crossover.   We finally   illustrate  these features in the context of  free standing graphene physics.
\end{abstract}

\maketitle 

\section{Introduction}

The discovery of  graphene  \cite{novoselov04,novoselov05}   has generated  an almost unprecedented      activity  devoted to  understanding   its   exceptional  mechanical, optical and  electronic  properties including high  electronic mobility, mechanical strength, optical transmittance  and  thermal  conductivity (see, e.g.,  \cite{katsnelson12} and references therein).   From  the  mechanical point of view  graphene appears,   due to its one-atom-thick layer structure and its elastic properties,  as the first  realization  of  a   perfect  two dimensional  (2D) crystalline  --  or polymerized -- membrane (see, e.g.,  \cite{castroneto09}).   From  the  electronic  point of view   graphene displays  a low-energy spectrum made of  massless chiral Dirac fermions  that mimic   quantum electrodynamics \cite{castroneto09}.   Also one of the most striking   features   of graphene and graphene-like materials is the     interplay  between   the   elastic and electronic   degrees of freedom.   For instance  elastic distortions   of  graphene   layers  modify   the electronic band structure  and  thus  the  conductivity  of this material \cite{katsnelson08,mariani08,guinea08b,kim08,castroneto09,vozmediano10,castro10}. Reciprocally  electronic degrees of freedom  have been proposed to be at the origin of   ripples  formation \cite{gazit09b,sanjose11,guinea14,gonzalez14}.    This  situation has opened   a very   promising road   for   the   design    of  graphene   and graphene-like materials  with   optimized optical and electronic  properties by  means of "strain engineering"    (see  \cite{roldan15}  and references therein).  From a more fundamental point of view  the specific  interaction between the   elastic and electronic degrees of freedom in graphene has been at the origin of   a very challenging   many-body problem combining  both high-energy and condensed matter  physics concepts that   has notably  received   an elegant and fruitful  formalization   in terms  of gauge-field theory and   field theory  -- Dirac equation -- in curved space  (see \cite{vozmediano10,amorim16} for reviews).   While  much  progress has  been  realized  toward an understanding of  the full problem combining  electronic and elastic degrees of freedom,  a complete and  quantitatively  well controlled description of the mechanical part of the problem,   that   is already associated with   a non-trivial interacting field theory, is still lacking.  

  From this point of view the  study  of    the joint  effects  of interactions and  fluctuations  on the long distance  properties of 2D  {\it classical} polymerized membranes  is already a well explored subject   (see \cite{proceedings89} for a  state of the art).   As revealed by early    studies  \cite{nelson87,aronovitz88,guitter89,aronovitz89,ledoussal92} the most important  physical effect taking place in  these systems    is   the stabilization of  long-range {\it orientational}  order  through the anharmonic coupling between stretching and  bending degrees of freedom.   Indeed the  coupling  between  the corresponding  in-plane, $\bf  u$,  and out-of-plane,  $\bf h$, modes  leads to a drastic reduction of    the fluctuations of  the latter  that are  otherwise, {\it i.e.} at the harmonic level,  divergent.  Technically   this can be understood through   the appearance of a  momentum dependent   effective  bending rigidity modulus $\kappa(q)\sim q^{-\eta}$ \cite{nelson87}, where  $\eta$ is    a positive exponent,   whose increase    at low momenta  enforces  the rigidity and thus  the stability of the membrane.  The  exponent  $\eta$ has been computed  by means of   analytical computations on the continuous effective theory of polymerized membranes  and by means of   Monte Carlo and molecular dynamics simulations  on discretized versions  of  membranes or on  models  of graphene  using empirical potentials (see \cite{amorim16} for a review).  Although a  significant dispersion of the results is observed   the  agreement between  the  analytical  predictions --   $\eta\in [0.789-0.85]$ --   and numerical computations  --  $\eta\in[0.72-0.85]$  --  leaves  little  doubt as for the adequacy  of continuous models of polymerized membranes to describe the  long distance  physics  of membranes as  well as   the elastic degrees of  freedom of   graphene (see for instance \cite{katsnelson13}).

Being given the  importance of  the  interplay  between anharmonicities and  {\it thermal}  fluctuations    regarding the question of stability  and, more generally, the  long distance behaviour,   of  membranes,  a  question that naturally arises   is that  of   the joint  role of   anharmonicities and {\it quantum}  fluctuations   in these systems.  This question is not purely formal since simple estimate  \cite{amorim14}   as well as   computations  by means of   a self-consistent field theoretical approach using a interatomic potential specific to graphene \cite{dasilva14},  have shown that, for this material,   the  typical temperature up to which quantum effects are significant  should lie around   $T\sim 1000$ K.   Several  recent works  have  been  devoted to  understanding the effects induced by quantum fluctuations  on membranes (see \cite{amorim16}  for  a review). In  \cite{sanjose11,guinea14,gonzalez14}   studies  including both  anharmonic elastic contributions  and  electron-phonon coupling  have  been performed using the  Self-Consistent Screening Approximation (SCSA),   extending the approaches  performed for  classical membranes  \cite{ledoussal92,gazit09,zakharchenko10}.  In these works it has been shown that  the   presence  of a strong  coupling between electronic  and elastic degrees of freedom  leads  to a drastic decrease of the bending rigidity  inducing  an  instability in the flexural sector. From the Renormalization Group (RG) point of view this instability  is reflected    in  the appearance of a critical point characterizing  a rippling transition.    In  the opposite  limit of vanishing electron-phonon  coupling,   the   RG  equations  for the Young  modulus  \cite{sanjose11,guinea14}  and bending rigidity modulus  \cite{sanjose11}   describe    a  (logarithmic) decrease  of the former and  a  (logarithmic)   increase of the latter,  leading to a stable  configuration for the membrane,  as in the classical case.  In \cite{kats14}  a    RG  approach of quantum membranes  has been  performed   in a perturbative context,  leading to  the opposite -- and surprising -- result that quantum fluctuations  induce  the  phenomenon   of  ultra-violet  asymptotic freedom, and thus a strong coupling  behaviour  in the infra-red.   According to   \cite{kats14}  this would lead   to a destabilization of the flat phase  at long distances.   Finally in   \cite{amorim14}  the thermodynamics  of  membranes at  very  low temperatures  has  been explored  through a  computation  of the self-energy  of the flexural mode  at one-loop and   then  in  a  self-consistent way     in the spirit of the early approach to  classical membranes of  Nelson and Peliti  \cite{nelson87}. The aim of   this approach,   that we  refer to as  Self-Consistent Born Approximation (SCBA) --   to be distinguished  from the SCSA by the fact that  only the flexural mode propagator, and thus not the interaction,  is treated  self-consistently --  was to   take   into account  both anharmonic and retardation interaction terms  neglected in previous  approaches \cite{guinea14,kats14}.  The  main  and, again, surprising  result   of this  approach is  the existence, at vanishing temperature,  of  a  {\it power-law} momentum scaling of  the bending rigidity modulus $\kappa(q)\sim q^{-\eta}$ with a critical exponent  $\eta=2$,  as opposed to  the  logarithmic behaviour obtained in previous approaches \cite{sanjose11,guinea14,gonzalez14,kats14}.  This   result  has been  the subject of a controversy   as for the interpretation of the corrections   to the self-energy obtained in this work (see \cite{kats14b,amorim14b} for  details).

 In this article we  develop an  approach   to quantum polymerized membranes   realized within  a   Nonperturbative Renormalization Group (NPRG)   framework, the effective average action method  \cite{wetterich93,berges02,delamotte03,pawlowski07,kopietz10,gies12,delamotte12,rosten12,nagy14},   extending previous works  performed  on classical membranes \cite{kownacki09,braghin10,essafi11,hasselmann11,essafi14}  to their  quantum counterpart.   We first derive the   general RG  equations   for a  model  of membranes  that  describes both  the phonon and flexural  modes and encompasses   both quantum and classical, {\it i.e.}  thermal, fluctuations.  We then focus on the flat phase  of membranes whose RG equations are obtained from the latter  via the  phenomenon of decoupling of  phonon modes.   At vanishing  temperature,    the   analysis of   the   equations  shows that  the  long distance behaviour  of membranes    is   controlled by a trivial -- gaussian --  fixed point  in  agreement with previous work performed  within  the SCSA framework    \cite{sanjose11,guinea14,gonzalez14}  but   in disagreement with the results  found within a recent perturbative approach in which an ultra-violet asymptotic freedom is predicted \cite{kats14}  and with those obtained within a recent  SCBA approach leading to a non-trivial critical exponent $\eta$  \cite{amorim14}.    At  any finite  temperature  the RG  flow drives the system  toward  an effective  high temperature  regime   which is well described by the classical  theory of membranes that predicts a  long distance behaviour governed by a non-trivial  infra-red fixed point.   This  crossover between the  short distance -- quantum -- regime and long-distance -- classical --  regime  as well as the crossovers    between weak and strong coupling  regimes  are  discussed within  the context of  free standing graphene physics.

Our article is organized as follows. In  Section II   we  present   the action    used to investigate   the long distance  behaviour of quantum polymerized membranes and discuss  the fluctuations  around the flat phase.  In section III  we describe   the   effective average action method  employed  to derive the NPRG equations for quantum membranes.  In section IV  we derive  the RG flow for  general membranes and, then,  derive and analyze that relevant to study their flat phase,  at finite,  at vanishing and at high temperatures,  and compare our results with previous works. In section V  we illustrate   our findings in the  context of the physics of free standing graphene.  Finally, in section VI, we conclude.

\section{The action  and   fluctuations around  the flat phase}

\subsection{The action for quantum membranes}

We  consider a   $D$-dimensional quantum membrane  embedded in a  $d$-dimensional Euclidean space.  Each point of the membrane  is parametrized by $D$ {\it internal}   coordinates ${\bf x}\equiv x_{\gamma}$,  $\gamma=1\dots  D$. Its location in the Euclidean space is realized 
through the embedding:  ${\bf x}\to {\bf R} ({\bf x},\tau)$ where  ${\bf R}$ is a $d$-component vector field with  components  $R_i, i=1\dots  d$ and $\tau$ an imaginary time (see, e.g., \cite{altland10}).  The action  of the model we consider is given by: 
\begin{align}
\displaystyle  S \left[{\bf R}\right]&=\displaystyle \int_0^\beta d\tau\:\int d^Dx\: \bigg\{{\frac{\rho}{2}(\partial_{\tau} {\bf R})^2}+\frac{\kappa}{2}(\partial_{\gamma}\partial_{\gamma}{\bf R})^2  \nonumber \\
&\hspace{-0.5cm}  \displaystyle+\tilde {\mu\over 4}(\partial_{\gamma}{\bf R}.\partial_{\nu}{\bf R}-\tilde\zeta^2\delta_{\gamma,\nu})^2+\tilde {\lambda\over 8}(\partial_{\gamma}{\bf R}.\partial_{\gamma}{\bf R}-D\tilde\zeta^2)^2\bigg\} 
\label{action}
\end{align}
where  $\tau$  takes  its  values   in $[0,\beta]$ with   $\beta=1/T$   ($\hbar$ and $k_B$ are taken equal to 1).   In Eq.(\ref{action}) 
the first term represents the kinetic part of the action,  the second one  a  bending energy  term while the other ones correspond   to harmonic and anharmonic elastic terms.   The stability of the action implies that   $\tilde \mu$ and  the  bulk  modulus $2\tilde \mu/D+\tilde \lambda$  must be positive.  The  action (\ref{action})  is the extension  to the quantum case of the action used to study classical polymerized membranes.  At mean-field level,  varying the  squared extension  factor $\tilde\zeta^2$,     it    describes a phase transition  between a crumpled phase for  $\tilde\zeta^2<0$,   with  a vanishing average value of   the tangent vectors $\langle \partial_{\gamma} {\bf R}\rangle$,  and a flat phase for  $\tilde\zeta^2>0$   defined  by:
\begin{equation}
\langle {\bf R}({\bf x},\tau)\rangle= \tilde\zeta\sum_{\alpha=1} ^D x_{\alpha}\:\bf{e}_{\alpha}
 \label{flatphase}
 \end{equation}
that  implies  $\langle \partial_{\gamma} {\bf R} \rangle=\tilde\zeta {\bf e}_{\gamma}$ with  $\gamma=1\dots  D$,  the ${\bf e}_{\gamma}$'s forming  an orthonormal set  of $D$ vectors that  generate  the flat   configuration (\ref{flatphase}).

The form  (\ref{action}), in which   the action is expanded around the  flat  configuration  (\ref{flatphase}),   is particularly  well-suited   to study  the flat phase of membranes  we are interested in.  It  is, in particular,  reminiscent of the action used to study {\it perturbatively}   this   phase  if one  replaces,   in Eq.(\ref{action}),  the strain tensor  $u_{\gamma\nu}=\partial_{\gamma}{\bf R}.\partial_{\nu}{\bf R}-\tilde\zeta^2\delta_{\gamma,\nu}$ by its expansion 
in terms of  in-plane   phonon  modes  $u_{\gamma}$  and  out-of-plane   flexural   modes ${\bf h}$: 
$u_{\gamma\nu}= {1\over 2}(\partial_{\gamma} u_{\nu}+\partial_{\nu} u_{\gamma}+\partial_{\gamma}{\bf h}.\partial_{\nu}{\bf h}+ \partial_{\gamma} u_{\sigma}\partial_{\nu} u_{\sigma}$). 
However we emphasize  the fact  that, within our approach, we do not   use  this  explicit  decomposition  in terms of  $\bf u$ and $\bf h$  fields.  Indeed  we  keep a full rotationally invariant formalism  during the whole computation where only the field ${\bf R}$ --  and its derivatives  -- come into play.   This is  possible  since  our approach being nonperturbative,  we are  not restricted  to study   weak coupling constants,    weak  fluctuations regimes. One is able, in particular,    to investigate  both the crumpling-to-flat transition and the flat phase using the same action, as done in \cite{kownacki09,braghin10,essafi11,hasselmann11,essafi14},  while the corresponding  fixed points are nonperturbatively connected  to each other.  Technically this implies, for instance,   that one  can, and must,  keep  the quadratic term  in the phonon field  $\bf u$ in the expression of $u_{\gamma\nu}$ above  that is  generally  discarded.

\subsection{Spectrum and fluctuations}
\label{spectrum}

One  can compute, from Eq.(\ref{action}),  the  propagator, and thus the spectrum of excitations  in  the flat phase.   The propagator   is given by  the second derivative  of   $S[{\bf R}]$  with respect to the  field ${\bf R}$, taken in the configuration Eq.(\ref{flatphase}).  In Fourier space  the  (imaginary) time direction is   compact  and, choosing periodic boundary conditions  in   time ${\bf R}({\bf x},\tau=0)={\bf R}({\bf x},\tau=\beta)$,    the field ${\bf R}(\bf x,\tau)$ can be expanded according to:
\begin{equation}
{\bf R}({\bf x},\tau)={1\over \beta}  \sum_{\omega_n} \int_q\   {\bf R}({\bf q},\omega_n) \ e^{i({\bf q}.{\bf x}-\omega_n \tau)}
\end{equation}
where  $\int_q$ stands for $\int d^Dq/(2\pi)^D$  and   the  $\omega_n$ are  Matsubara  frequencies: $\omega_n=2\pi n/ \beta$, $n\in \nbZ$. 
The  resulting propagator  splits  into three  parts: 
\begin{subequations}
\begin{empheq}[left=\empheqlbrace]{align}
&G_F^{-1}({\bf q},\omega_n)=\kappa  {q}^4+\rho \omega_n^2 \label{propagata}\\
\nonumber \\
&G_{\parallel}^{-1}({\bf  q},\omega_n)=\kappa ({q}^4+\widetilde{m}_2^2\,  {q}^2)+ \rho \omega_n^2 \label{propagatb}\\
\nonumber\\
&G_{\perp}^{-1}({\bf  q},\omega_n)=\kappa ({q}^4+\widetilde{m}_1^2  {q}^2) +\rho \omega_n^2 \label{propagatc}
\end{empheq}
\end{subequations}
with $\widetilde{m}_1^2= \tilde \mu   \tilde\zeta^2 \kappa^{-1}$ and  $\widetilde{m}_2^2=(2 \tilde\mu+\tilde\lambda)\tilde\zeta^2 \kappa^{-1}$ and  where $q=\vert \bf q\vert$.   In Eqs.(\ref{propagata})-(\ref{propagatc}),   $G_F^{-1}({\bf  q},\omega_n)$  is associated with   $d-D$ flexural   modes;    $G_{\perp}^{-1}({\bf q},\omega_n)$   and $G_{\parallel}^{-1}({\bf  q},\omega_n)$  are  associated respectively  with  the   $D-1$   transverse (directed orthogonally to  ${\bf q}$) and a single  longitudinal  (directed along ${\bf q}$) phonon modes. We  have also  introduced  two squared masses $\widetilde{m}_1^2$    and $\widetilde{m}_2^2$  in the propagators  of the transverse and longitudinal phonon modes.    This rather unusual  way to represent  these latter modes   as  a kind  of "massive"  flexural modes  will take its full meaning  in the context of the NPRG approach below where masses  acquire a dynamical, running, status.

 The  expressions (\ref{propagata})-(\ref{propagatc})  can be used to evaluate the importance of fluctuations in the flat phase.    Due to the  dispersion relation of flexural modes  it   appears   that the fluctuations  associated with  these modes  are  most important; hence  we concentrate on them. 
At high temperatures   only the  vanishing  frequency mode contributes and the propagator  of  the flexural --  $\bf h$ --   modes, at low momenta,  is given by the classical expression:
\begin{equation}
G_F({\bf q},\omega_n) \underset{q\to 0, T\to\infty}{\sim}  {1\over \kappa  q^4}\  .
\end{equation}
This allows  to compute the fluctuations of the normals  to the membrane that characterize the  existence or absence of long-range orientational order. Standard computations  show   that  they are essentially given by the  spatial  average  of  the correlation function of   the $\partial_{\gamma}  {\bf h}$  field. One finds, at the harmonic   level:
\begin{equation}
 \langle\langle (\partial_{\gamma} {\bf h})^2\rangle\rangle_{h}  \sim  \int \ d^D q\  q^2\, {1\over \kappa  q^4} \sim L^{2-D} 
\end{equation}
where $L$ is the typical size of the system and   $\langle\langle\dots\rangle\rangle$ denotes both a  thermodynamical and  momentum  average.  One recovers the  classical result   that   the system displays  quasi-long range orientational order in two dimensions. This  behaviour is modified by anharmonic contributions that induce  a $q$-dependence of  the bending  rigidity modulus, $\kappa(q)\sim  q^{-\eta}$ \cite{nelson87},  and   make the integral 
\begin{equation}
 \langle\langle (\partial_{\gamma} {\bf h})^2\rangle\rangle_{anh} \sim \int \ d^D q\  q^2\ {1\over \kappa(q)\,   q^4} \sim L^{2-D-\eta} 
\end{equation} 
convergent   in $D=2$ for any positive value of $\eta$. 

At very low temperatures, all Matsubara frequencies contribute  and  they  can be considered as continuous. One thus has:
 \begin{equation}
G_F({\bf q},\omega_n) \underset{q\to 0, T\to 0}{\sim}  {1\over \kappa  q^4+\rho \omega^2}
\end{equation}
so that : 
\begin{equation}
\langle\langle (\partial_{\gamma}  {\bf h})^2\rangle\rangle_{h}  \sim \int \ d^D q\,  d\omega\, \ {q^2\over \kappa  q^4+\rho \omega^2} \sim L^{-D} 
\end{equation}
 which  converges in particular  in $D=2$.  This means that quantum fluctuations  are not strong enough to destabilize the long-range orientational order unless   interactions   change  this behaviour.

 \section{The effective average action method}
 
 We now present   the  effective average action method  that we  use to compute the effects of   interactions   in quantum membranes.

\subsection{General principle}

 The  effective average action method    (see  \cite{wetterich93} and \cite{berges02,delamotte03,pawlowski07,kopietz10,gies12,delamotte12,rosten12,nagy14} for reviews)    achieves, at the level of  the (Gibbs) free energy, the Wilson-Kadanoff bloc-spin  program.  It  involves  a central object,   $\Gamma_k$, --  $k$ being a running scale -- that  is   essentially   a  running Gibbs free energy where only fluctuations of momenta $q\ge k$  have been integrated out.   Note that we consider,   for the time being, a {\it classical}  situation and thus   a {\it classical}  free energy.   At the microscopic lattice  scale  $\Lambda$, no  fluctuation  has been integrated out and $\Gamma_{k=\Lambda}$  identifies with  the microscopic action  $S$ while,   at long distances, { \it i.e.} at $k=0$, it identifies with the  standard Gibbs free energy $\Gamma$: 
\begin{subequations}
\begin{empheq}[left=\empheqlbrace]{align}
\Gamma_{k=\Lambda}=S\   \label{limitesgammaa} \\ 
\nonumber \\
\Gamma_{k=0}=\Gamma \label{limitesgammab}\  
\end{empheq}
\end{subequations}
and, at  any  finite momentum  scale $0<k<\Lambda$,  $\Gamma_k$  interpolates smoothly between 
these two limits.      In practice   the integration over the  high momenta --  $q>k$ -- modes  is realized by  adding to the microscopic action, Eq.(\ref{action}),    a  $k$-dependent  "mass" term : 
\begin{equation}
\Delta S_k[{\bf R}]={1\over 2} \int_{\bf q} \, R_{k,ab}({\bf q})\, R_a ({\bf q}) R_b(-{\bf q}) 
\label{massterm}
\end{equation}
 so that the partition function,   in  presence of a source term,  reads:
 \begin{equation}
 {\cal Z}_k[{\bf J}]=\hspace{-0.1cm}\int {\cal D}{\bf R} \exp\Big(-S[{\bf R}]-\Delta S_k[{\bf R}]+ \int_{\bf q}  {\bf J}({\bf q}).{\bf R}(-{\bf q})\Big) \  .
 \label{partition}
\end{equation}
Note that,   in the rest of the article,   the function $R_{k,ab}({\bf  q})$,  discussed below,  will be considered as diagonal in field space, {\it i.e.}  $R_{k,ab}({\bf  q})=R_k({\bf q})\delta_{a,b}$. 

The role of the  $\Delta S_k[{\bf R}]$ term in Eq.(\ref{partition})  is to "ballast" the low momenta  modes and to make that only the high momenta modes are effectively integrated out.  To realize this program, and  to make  sure   that   $\Gamma_k$    meets   the conditions  Eq.(\ref{limitesgammaa})-(\ref{limitesgammab}),   the function $R_k({\bf  q})$  must obey  several constraints:      {\it i)}    it behaves as  $k^{\alpha}$   -- with  suitable  power $\alpha>0$ -- at low momenta  {\it ii)}   it vanishes at   high   momenta:
\begin{subequations}
\begin{empheq}[left=\empheqlbrace]{align}
&R_k({\bf  q})\sim k^{\alpha} \hspace{0.5cm} \hbox{when} \hspace{0.5cm}  q\ll k   \label{relationR1a} \\
\nonumber \\
&R_k({\bf  q})\to 0 \hspace{0.6cm} \hbox{when} \hspace{0.5cm} q\gg k  \label{relationR1b} \ .
\end{empheq}
\end{subequations}

The  relation  (\ref{relationR1a})  implies that the low momenta  modes are affected  by a "mass"  term Eq.(\ref{massterm}) with mass $k^{\alpha}$ that prevents  their  propagation  -- they "decouple" -- while the  relation (\ref{relationR1b})  implies that the high momenta  modes are kept untouched.  These two relations  can be  translated  in terms of the $k$-dependence of $R_k({\bf  q})$: 
\begin{subequations}
\begin{empheq}[left=\empheqlbrace]{align}
&R_k({ \bf  q}) \sim \Lambda^{\alpha} \hspace{0.5cm} \hbox{when} \hspace{0.5cm}  {k\to\Lambda}  \label{relationR2a}\\
\nonumber \\
&R_k({\bf  q})\to 0  \hspace{0.7cm} \hbox{when} \hspace{0.5cm}  {k\to 0} \label{relationR2b} \ . 
\end{empheq}
\end{subequations}

The   relation (\ref{relationR2a})  means  that,  when $k\to \Lambda$,  $\Delta S_k[{\bf R}]$ acts as  a   large mass term for all modes, what  implies  that  almost no mode contributes to the functional integral in  Eq.(\ref{partition});  thus  no  fluctuation  has  been  included. The relation (\ref{relationR2b})  means  that,  when $k\to 0$,  $\Delta S_k[{\bf R}]$   plays no  role,   so   that  all   modes contribute  to the functional integration;  thus all fluctuations have been included.  A typical cut-off function  satisfying all the previous requirements   is given by the "$\Theta$"  cut-off  \cite{litim00}: 
\begin{equation}
R_k({\bf  q})=Z_k  q^{\alpha}\Big(\Big({k\over q}\Big)^{\alpha}-1\Big) \Theta\Big(1-{q^2\over k^2}\Big)
\label{cutoff1}
\end{equation}
where  $\Theta$ is the  step function and  $Z_k$ is a  field renormalization -- see below.  Note that the extension to a quantum  system can  lead  to consider  $R_k$ as a function of both momentum ${\bf  q}$ and frequency  $\omega$ -- see below. We now define  the  running Gibbs free energy $\Gamma_k$ as a  modified  Legendre transform of  the Helmholtz free energy  $W_k[{\bf J}]=\ln {\cal Z}_k[{\bf J}]$  \cite{berges02,delamotte03,pawlowski07,kopietz10,gies12,delamotte12,rosten12,nagy14}:
\begin{equation}
\Gamma_k[{\bf r}]=-W_k[{\bf J}]+ \int_{\bf q}  {\bf J}({\bf q}).\,{\bf r}(-{\bf q}) -\Delta S_k[{\bf r}]
\label{defgamma}
\end{equation} 
where   ${\bf r}$  is the  expectation value of the microscopic field ${\bf R}$   in presence of  an external source ${\bf J} $:
\begin{equation}
r_i({\bf x})=\langle  R_i({\bf x})\rangle={\delta W_k[{\bf J}]\over \delta  J_i({\bf x})}
\end{equation}
and  $\Delta S_k[{\bf r}]$ is   the  macroscopic  counterpart of the microscopic  mass term Eq.(\ref{massterm}).

From   Eqs.(\ref{defgamma}) and (\ref{relationR2b})  one sees that,  when  $k\to 0$,   $\Gamma_k[{\bf r}]$ coincides with  the usual free energy  $\Gamma$.   The  opposite limit $k\to \Lambda$  is a little bit more tricky. From Eq.(\ref{defgamma}) one has: 
\begin{equation}
 J_i({\bf q})={\delta\Gamma_k[{\bf r}]\over \delta{r_i}(-{\bf q})}+  R_k({\bf q})\, r_i({\bf q})\ . 
\label{J}
\end{equation} 
Using then  Eqs.(\ref{partition}), (\ref{defgamma}) and (\ref{J}) one gets:
\begin{equation}
\begin{split}
e^{-\displaystyle \Gamma_k[{\bf r}]}=&\int \mathcal D{\bf R}\ \exp\Big(
-S[{\bf R}]\\
&+ \int_{\bf q} {\delta \Gamma_k[{\bf r}]\over \delta r_i({\bf q})} \big(R_i({\bf q})-r_i({\bf q})\big)-\Delta
S_k[{\bf R}-{\bf r}]\Big)\ .
\label{fonctionnal2}
\end{split}
\end{equation}
In the limit $k\to \Lambda$, $R_k({\bf q})$ is, from Eq.(\ref{relationR2a}),  very large  -- infinite if we assume $\Lambda\to \infty$  --  so that  the mass term   ${\hbox{exp}}(-\Delta S_k[{\bf R}-{\bf r}])$  acts essentially as a hard constraint,   $\delta ({\bf R}-{\bf r})$,  in the functional integral  Eq.(\ref{fonctionnal2}) \footnote{Cut-off functions  $R_k({\bf q})$  that  effectively diverge  in the limit $k\to \Lambda$, and thus that lead  rigorously to a   $\delta ({\bf R}-{\bf r})$ constraint,   can be easily  constructed.}. In this way  $\Gamma_{k=\Lambda}[{\bf r}]$   identifies with the microscopic action  $S[{\bf r}]$,   Eq.(\ref{action}). We thus  fully recover  the constraints of  Eqs.(\ref{limitesgammaa})-(\ref{limitesgammab}). \\

\subsection{The Wetterich  equation}

Once the limits of  $\Gamma_k$  have  been given it remains to precise its $k$-dependence.   It   is provided   by  the Wetterich equation \cite{wetterich93c}: 
\begin{equation}
\partial_t \Gamma_k[{\bf r}]={1\over 2} \hbox{Tr} \Big\{{\partial_t  R_k}\,  (\Gamma_k^{(2)}[{\bf r}]+R_k)^{-1}\Big\}
 \label{renorm}
\end{equation}
where one defines a RG "time"  $t=\ln \displaystyle {k / \Lambda}$ and where  the trace has to be understood as   a  space (or momentum)  $D$-dimensional integral   as well as a summation over internal indices.   An equivalent and useful  expression is given by:
\begin{equation}
\partial_t \Gamma_k[{\bf r}]={1\over 2} {\widehat{\partial_t}} \hbox{Tr} \ln \big(\Gamma_k^{(2)}[{\bf r}]+R_k\big)
 \label{renormbis}
\end{equation}
where $\widehat{\partial}_t$ means that  the $t$-derivative  only acts on $R_{k}$. 
 In Eqs.(\ref{renorm}) and (\ref{renormbis})  $\Gamma_k^{(2)}[{\bf r}]$  is the  inverse propagator, the second derivative of $\Gamma_k$ with respect to the field $\bf r$:
\begin{equation}
\Gamma_{k,ij}^{(2)}[{\bf r};{\bf q},{\bf -q}]={\delta^2 \Gamma_k[{\bf r}]\over\delta r_i({\bf q})\delta r_j({\bf -q})}
\label{propag}
\end{equation}
 taken in a {\sl generic}  field configuration.  Physically Eq.(\ref{renorm}) describes  the evolution of  the effective action $\Gamma_k$ when $k$ is lowered and when fluctuations at lower and  lower momentum scales  are  taken into account, its dynamics being controlled by  the $k$-dependence of $R_k$, $\partial_t  R_k$.    The cut-off  function  $R_{k}$  also enters   Eq.(\ref{renorm})  as an additive term to   the inverse propagator  $\Gamma_k^{(2)}$ and  directly impacts   the propagation  of the corresponding  mode of momentum $q$ in the  implicit integral in Eq.(\ref{renorm}), what allows to realize  the  bloc spin program described above.  
 
 Technically two properties of $R_k$ should be emphasized. First, since   $R_k({\bf  q})\sim  k^{\alpha}$  at  low momenta it acts as  an  effective  infra-red regulator  for the corresponding modes and   prevents the occurrence of  infra-red divergences  at any finite value of $k$.    The exponent  $\alpha$ is  chosen to realize this aim; it is generally taken equal to $2$ in  usual,  $O(N)$-like, theories, that   involve a spatial  kinetic term of order $\partial^2$. In the  case of membranes  it is taken equal to 4  since, there,   the   spatial  kinetic  term is  of order  $\partial^4$. The  regular behaviour that follows  allows to  investigate the critical physics by progressively lowering $k$ without having recourse to  an $\epsilon$-expansion or other  similar  tools.  Second,  since $\partial_t R_k({\bf  q})$ contributes  essentially  for    momenta $q\le k$, as it can be checked  directly using the expression Eq.(\ref{cutoff1}), it  makes the RG flow ultra-violet finite.  

\subsection{Properties}
\label{properties}

We  now discuss the main  properties of  Eq.(\ref{renorm}), which  have   been otherwise  extensively described in several reviews \cite{berges02,delamotte03,pawlowski07,kopietz10,gies12,delamotte12,rosten12,nagy14}.   First Eq.(\ref{renorm}) is an  {\it exact} equation. It  encompasses  all perturbative  (magnons, phonons, flexurons modes, etc) and nonperturbative  (bound states, topological excitations, instantons, etc) features   of the  underlying theory.   Second   Eq.(\ref{renorm})   involves  a single momentum  integral and  thus displays   a {\it one-loop}  structure.   This makes  the  momentum dependence of  computations  much  simpler  than their perturbative counterparts.  In practice this  one-loop structure allows a direct comparison  with the leading order  of all  perturbative computations: weak-coupling, low-temperature, $1/N$ or $1/d$ expansions, etc.  As an illustration replacing  in Eq.(\ref{renorm})  $\Gamma_k^{(2)}[{\bf r}]$  by the second derivative of the classical action, $S^{(2)}$,  leads to the usual  one-loop effective action \cite{berges02}: 
\begin{equation}
\Gamma_k[{\bf r}]=S[{\bf r}]+{1\over 2} {\hbox{Tr}}\ln (S^{(2)}[{\bf r}]+R_k)\ . 
\label{gammaoneloop}
\end{equation}

Of course the complexity  is hidden elsewhere, precisely in the fact that Eq.(\ref{renorm}) involves the full field-dependent propagator $\Gamma_{k}^{(2)}[{\bf r}]$.    This  situation fundamentally differs   from  that  met within the   perturbative framework  were   the propagator  is generally considered  in a {\it vanishing}  field configuration.   This is, of course, a crucial  point of this framework  that   allows for   nonperturbative investigations.  Nevertheless this complexity forces us to perform approximations that   mainly consist  in doing  truncations of the effective action  $\Gamma_k[{\bf r}]$.   In this way  the computations are   doable  and  the  nonperturbative character of the method  is kept  intact  as far as the right-hand side of Eq.(\ref{renorm}) is  not expanded in powers of  any usual  small  parameter as a coupling constant, the temperature, $1/N$, $1/d$ etc.  Improving the  {\it ansatz}  allows to check the stability of the results  and then the adequacy  of the  truncation used. 
 
 An efficient truncation scheme   (see  \cite{berges02,delamotte03,pawlowski07,kopietz10,gies12,delamotte12,rosten12,nagy14}  for discussions concerning truncations) is the derivative expansion  where $\Gamma_k$ is expanded in  powers of the derivatives  of the order parameter.  It is justified when one focuses  on the critical physics, or  more generally, on the long distance behaviour  of a system and also when this  behaviour is controlled by the elementary excitations as opposed to the case where bound states occur.    In this last case, as in other situations involving for instance fermionic excitations,  one  has  to  take into account  the whole momentum structure and other kinds of  approximations should  be used (see for instance \cite{blaizot06,blaizot06b,blaizot06c,kopietz10}).  The derivative expansion  can be   -- and is often -- combined  with an  expansion  in powers of the order parameter  itself  around a given field configuration,  generally the minimum of the effective action.   The   great benefit of  this  combined derivative/field expansion is  to convert the functional Eq.(\ref{renorm}) to a set of differential equations for the coupling constants entering in the ansatz, as in the perturbative context.  Reaching  accurate   results can require  to deal with high  orders, see for instance \cite{canet03b}.      However  we insist on the fact  that, even at the lowest orders,    this expansion  provides qualitatively  -- and also  sometimes  quantitatively -- correct results in  numerous   contexts (see \cite{berges02,delamotte03,pawlowski07,kopietz10,gies12,delamotte12,rosten12,nagy14}).  This relies on the very structure of  Eq.(\ref{renorm}) that, even  approximated by means of  truncations of the action,  remains  nonperturbative.

\subsection{Effective average action for quantum membranes}
  
 In the present work  we  consider a truncation directly  inspired by  Eq.(\ref{action}) that reads: 
\begin{align}
\displaystyle  \Gamma_k\left[{\bf r}\right]&=\displaystyle \int_0^\beta d\tau\:\int d^Dx\: \bigg\{{\frac{Z_k^{\tau}}{2}(\partial_{\tau} {\bf r})^2}+\frac{Z_k}{2}(\partial_{\gamma}\partial_{\gamma}{\bf r})^2 \nonumber \\
\label{action2}  
&\hspace{-0.5cm}  \displaystyle+{\mu_k\over 4}(\partial_{\gamma}{\bf r}.\partial_{\nu}{\bf r}-\zeta_k^2\delta_{\gamma,\nu})^2+{\lambda_k\over 8}(\partial_{\gamma}{\bf r}.\partial_{\gamma}{\bf r}-D\zeta_k^2)^2\bigg\}  \ . 
\end{align}
This expression is deduced  from  the action (\ref{action})  by  replacing the microscopic field ${\bf R}$  by the macroscopic one   ${\bf r}$,  followed by the    usual  rescaling of the field:  ${\bf r}\mapsto  Z_k^{1/2} \kappa^{-1/2} {\bf r}$. We have  introduced    two field renormalization  "constants":  $Z_k$  and  {\bf $Z_k^{\tau}=Z_k \rho \kappa^{-1}$} and  defined  two elastic running coupling constants   $\lambda_k=Z_k^2\, \tilde \lambda\,   \kappa^{-2}$ and  $\mu_k=Z_k^2\,  \tilde \mu\,   \kappa^{-2}$ as well as   a  running extension parameter  $\zeta_k=\kappa^{1/2}Z_k^{-1/2}\tilde\zeta$ that, together,   parametrize the RG flow.    The ansatz (\ref{action2}) corresponds  to the  next-to-leading order  \footnote{The leading order -- called  {\it local potential approximation} would consist to  fix the  field-renormalizations    $Z_k$ and $Z_k^{\tau}$ to  1  leading to vanishing anomalous dimensions.} of  the   derivative  expansion  of the effective action  whose next  orders    would enclose   higher powers of   time and space  derivatives of  the field ${\bf r}$.   {Our  choice  to  limit ourselves  to  the form  (\ref{action2})    is  justified   by  the   fact that we   give here  priority to the description of   infra-red   behaviour of  the flat phase,  where only low-order derivative  terms are expected to play a  significant role, leaving  the  investigation  of the short distance behaviour to further work \cite{coquand16}.   The ability  of our ansatz (\ref{action2}) to describe  the former  regime has received  justifications in recent works.  For instance     Braghin and Hasselmann \cite{braghin10,hasselmann11}  have   studied    the flat phase of classical membranes  by means  of a  NPRG  approach  taking into account  the full momentum dependence  of the interaction vertices of Eq.(\ref{action2}) -- thus involving   infinite  order derivative terms. They   have shown  that the  values  of physical quantities,  as  the critical exponent $\eta$ characterizing  the  momentum dependence of the bending rigidity in this phase,  were  not impacted  by   derivative terms beyond the order considered here.    Note, moreover,    that we  have also performed   an expansion of the effective action  in powers  of the order parameter fields   ${\bf \partial_{\gamma}  r}$  and   ${\bf \partial_{\tau}  r}$ at the lowest non-trivial order.  A  complete treatment   of the  next-to-leading order  of the derivative expansion considered  here would require to use   for  the "kinetic " part   of the action not just  field renormalization constants   $Z_k$  and  $Z_k^{\tau}$  but  full {\it functions}   $Z_k({\partial_{\tau} \bf r},{\partial_{\gamma} \bf r})$  and  $Z_k^{\tau}({\partial_{\tau} \bf r},{\partial_{\gamma} \bf r})$  and for the elastic -- potential -- part   of the action a  full  "elasticity" function $U_k({\partial_{\gamma}  \bf r})$.   Beyond its simplicity,  the justification of   our choice comes  from the fact  that, at least in the classical case,  the critical exponent  $\eta$  in the flat  phase  is  strictly  {\it independent }  of powers of the field  higher    than four \cite{essafi14}; one expects this  result to remain  true in the quantum case.   Being given the ansatz Eq.(\ref{action2}) its  RG flow is given by (see \cite{berges02} and references therein): 
\begin{equation}
\begin{array}{ll}
\partial_t\Gamma_k[{\bf r}] = \displaystyle \frac{T}{2}\sum_{\omega_n}\hspace{-0.1cm} \int_{\bf q} \partial_t R_k({\bf \tilde q}) \ \big({\Gamma^{(2)}_k[{\bf r};{\bf \tilde q},-{\bf \tilde q}]+R_k({\bf  \tilde q})} \big)^{-1}_{ii} 
\label{renorm2}
\end{array}
\end{equation}
where  Eq.(\ref{renorm}) has been adapted  to the quantum case \footnote{See for instance \cite{berges02} and references therein for general considerations about  the NPRG approach to quantum systems  and, for  instance,  \cite{rancon13} for an approach  of  the quantum relativistic  $O(N)$ case,  close in spirit to the present one.}  In this expression ${\bf \tilde q}$ stands for  ${\bf \tilde q}=({\bf q},\omega_n)$  and  the trace in Eq.(\ref{renorm}) has been  written  explicitly   as  an integral over momentum ${\bf q}$,  a sum over the  frequencies $\omega_n=2\pi n/\beta$ as well as an implicit sum  over the field indices $i$. 

\subsection{The cut-off function $R_k$}
\label{cutoffsection}

It remains to specify  the functional dependence of the cut-off function $R_k(\tilde {\bf  q})$.  We consider the following expression: 
\begin{equation}
R_k(\tilde{\bf  q})=Z_k \left(q^4+{\omega_n^2\over  \Delta_k^2}\right) {\cal R}\left({q^4+{\omega_n^2/  \Delta_k^2}\over k^4}\right)  
\label{cutoff2}
\end{equation}
where ${\cal R}(.)$ is a function such that   $R_k({\bf  q},\omega_n=0)$ fulfills   the conditions Eq.(\ref{relationR1a})-(\ref{relationR1b}).  In Eq.(\ref{cutoff2}) $\Delta_k^2=Z_k/Z_k^{\tau}$ that  equals $\Delta_{\Lambda}^2=\kappa/\rho$ at the lattice scale.  The function $R_k(\tilde{\bf  q})$ acts now as a infra-red  cut-off for both low-momenta  and low-frequencies.   With this form its  respects the dispersion relation of the flexural modes that dominate  the flat phase.  As a consequence    modes with momenta  ${q}^2<k^2$ and/or frequencies  $\vert \omega_n\vert < \Delta_k k^2$  are eliminated while those with  momenta  ${q}^2>k^2$ and/or frequencies $\vert \omega_n\vert  >\Delta_k k^2$ are left untouched.   

As for the question of the  choice of the function ${\cal R}$  entering in  Eq.(\ref{cutoff2})  it  is generally a complex  one.   Indeed   due to the vanishing  property of the cut-off function $R_k({\bf q})$ when $k\to 0$, the   physical quantities extracted from   Eq.(\ref{renorm}) in this limit  are {\it a priori}  cut-off independent.  This  statement is however only true when  the effective action $\Gamma_k$  is treated {\it exactly}.   Yet, as  said above, in practice  one has to resort to approximations  of the effective action that destroy  the cut-off independence of  $\Gamma_{k=0}=\Gamma$ as computed from Eq.(\ref{renorm}), and thus that of  any physical quantity.   This situation has lead to the  search for the best or  optimal   regulator that  would  provide  the weakest  cut-off  dependence (see for instance \cite{pawlowski07} for a review).   This  question of  the optimization of the cut-off is a delicate one since one should precise what should  be optimized:  stability  \cite{litim00,litim01,litim01b,litim02},  cut-off  independence  of the  RG flow  \cite{pawlowski07},  speed of convergence of the field and field/derivative expansion, accuracy of physical quantities  \cite{canet03a}, etc. At  the  lowest order of the derivative expansion, called Local Potential Approximation,  where the field renormalizations are neglected, these different criterions coincide and  have  lead to  pick out  the   $\Theta$ cut-off  Eq.(\ref{cutoff1}) as the optimal one.  However beyond order $\partial^2$ the $\Theta$  cut-off no longer regulates  the RG flow and no  equivalent optimal cut-off has been proposed yet (see however \cite{nandori13,marian14} for  optimized cut-offs  that both regulate the flow and encompass several cut-offs previously  considered). In the present work, since previous  computations  performed on  classical membranes have shown an extremely  weak dependence of the results with respect to the field/field derivative content of the action \cite{kownacki09,braghin10,hasselmann11}   we opt for the usual  $\Theta$ cut-off that displays  both  formal simplicity and easy computability. 

Note finally, in Eq.(\ref{cutoff2}),  the presence of the  field renormalization $Z_k$ in front of the cut-off function that ensures that  $Z_k$   explicitly  disappears from  the  RG equations where it only appears through its derivative $\eta_k=-(1/Z_k) \partial_t Z_k$ -- see below.  At a fixed point $\eta_k$ reaches a fixed value $\eta$. This provides the scaling of $Z_k$:  $Z_k\sim k^{-\eta}$.   In the same way  $Z_k^{\tau}$ scales  at a  fixed point  as $Z_k^{\tau}\sim k^{-\tilde\eta}$.   

\subsection{Power counting}
\label{power}

 Action (\ref{action2})   involves  the  terms  that are supposed to  play  the most    important role at long distances;  this  allows  power counting considerations  at a putative fixed point.     In terms of running momentum scale  $k$   one has     $[\partial_{\gamma}]=k$ and  $[\partial_{\tau}]=[\partial_{\gamma}]^z=k^z$ were we have   introduced   the  dynamical critical exponent $z$.  Using the fixed point relations  $Z_k\sim k^{-\eta}$ and $Z_k^{\tau}\sim k^{-\tilde\eta}$ one    gets    the field dimension $[{\bf r}]=D-4+\eta+z$ or indifferently  $[{\bf r}]=D+\tilde\eta-z$ from which one deduces   the relation: $\tilde \eta=\eta+2(z-2)$.  As for the  elastic  coupling constants  one has:  $[\lambda]=[\mu]=-(D-4+z+2 \eta)$ and for the (squared) extension factor $[\zeta^2]=D-2+\eta+z$.  Simple  dimensional analysis -- {\it i.e.} $\eta=\tilde\eta=0$ --  leads to $z=2$   and  $[\lambda]=[\mu]=-(D-2)$.  This   defines  $D=2$ as the upper critical dimension of the theory  (\ref{action2})  as it will be confirmed by our RG analysis.

\section{The RG flow}

 \subsection{General  principle}

 \label{Generalprinciple}

 We now derive the RG equations for the different coupling constants entering in Eq.(\ref{action2}).  In order  to do this one has to define these coupling constants as   functional derivatives  of $\Gamma_k$   taken in a specific field configuration. We choose a  flat  ground state configuration, given by  Eq.(\ref{flatphase}),  since it corresponds to that  with a vanishing external source, the situation that one wants to  describe precisely.  In Fourier space it reads:
\begin{equation}
r_{k,j}^f({\bf q},\omega_n)=-i\zeta_k\,\delta_{n,0}\,\delta_{\gamma, j}\ \frac{\partial}{\partial q_{\gamma}}\delta({\bf q})
\label{rmin}
\end{equation}
where we recall that $\gamma=1\dots D$ and $j=1\dots d$.  Note that the configuration ${\bf r}_{k}^f$  also runs with $k$ due to the running of the extension parameter $\zeta_k$. 
	
With the action Eq.(\ref{action2}) a  generic coupling constant $g_{k,ij}$ can be defined as  some  coefficient   in the expansion of  $\Gamma^{(2)}_{k,ij}[{\bf r};{\bf\tilde{p}},-{\bf\tilde{p}}]$,   with ${\bf\tilde{p}}=({\bf{p}},\omega_m)$,    in powers of   $p$ and   $\omega_m$  (the explicit  definition of  each coupling constant are given in  Appendix(\ref{coupling}), Eqs.(\ref{defmu})-(\ref{defztau})). Explicitly one has: 
\begin{equation}
g_{k,ij}=\lim_{\tilde{{\bf p}}\rightarrow{\bf 0}}\frac{1}{a!b!}\frac{d^a}{d(\omega_m^2)^a}\frac{d^b}{d(p^2)^b} \Big\{ \Gamma^{(2)}_{k,ij}[{\bf r};{\bf\tilde{p}},-{\bf\tilde{p}}]\big |_{{\bf r}_{k}^f} \Big\}\ . 
\end{equation}
The flow of  the  coupling constant $g_{k,ij}$ is then  deduced by taking a derivative with respect to $t$:
\begin{align}
\partial_t g_{k,ij}=&\lim_{\tilde{{\bf p}}\rightarrow{\bf 0}}\frac{1}{a!b!}\frac{d^a}{d(\omega_m^2)^a}\frac{d^b}{d(p^2)^b}  \Big\{\partial_t\Gamma^{(2)}_{k,ij}[{\bf r};{\bf\tilde{p}},-{\bf\tilde{p}}]\Big \vert_{{\bf r}_{k}^f}  \nonumber \\
 & +  \int_{\bf \tilde{ q}}\Gamma^{(3)}_{k,ijl}[{\bf r};{\bf\tilde{p}},-{\bf\tilde{p}},{\bf\tilde{q}}]\  \partial_t r_k^l ({\bf\tilde{q}})  \Big |_{{\bf r}_{k}^f}\Big\} \ . 
 \label{flowg}
\end{align}
It results from the previous equation  that it  is sufficient to know the flow of $\Gamma^{(2)}_{k,ij}[\mathbf{r};{\bf\tilde{p}},-{\bf\tilde{p}}]$ which is easily obtained by taking the second derivative of Eq.(\ref{renorm2}) with respect to $r_i({\bf\tilde{p}})$ and $r_j({\bf\tilde{p}})$, what leads  to:
\begin{widetext}
\begin{equation}
\begin{array}{ll}
\partial_t\Gamma^{(2)}_{k,ij}[{\bf r};{\bf\tilde{p}},-{\bf\tilde{p}}]= & \displaystyle    -\frac{1}{2}\widehat\partial_t\bigg\{
\displaystyle \int_{\bf\tilde{q}} \hspace{-0cm}G_{k,ab}[\mathbf{r},{\bf\tilde{q}}]\ \Gamma^{(4)}_{k,ijab}[\mathbf{r};{\bf\tilde{p}},-{\bf\tilde{p}},{\bf\tilde{q}},-{\bf\tilde{q}}]\\ 
& -\displaystyle  \int_{\bf\tilde{q}} {G}_{k,ab}[\mathbf{r},{\bf\tilde{q}}]\, \Gamma^{(3)}_{k,iac} [\mathbf{r};{\bf\tilde{p}},-{\bf\tilde{q}},{\bf\tilde{q}-\tilde{p}}] \   G_{k,cd}[\mathbf{r},{\bf\tilde{q}-\tilde{p}}]\ \Gamma^{(3)}_{k,jbd}[\mathbf{r};-{\bf\tilde{p}},{\bf\tilde{q}},{\bf\tilde{p}-\tilde{q}}]\bigg\}
\end{array}
\label{gamma2}
\end{equation}
\end{widetext}
where
\begin{equation}
{G}_{k,ab}[\mathbf{r},{\bf \tilde q}]=[\Gamma^{(2)}_{k,ab}[\mathbf{r}; \tilde{\bf q},-\tilde{\bf q}]+R_{k,ab}(\tilde{\bf q})]^{-1}\ . 
\end{equation}
In Eq.(\ref{flowg}) and (\ref{gamma2}) the vertices $\Gamma^{(3)}_{k,bjd}$ and $\Gamma^{(4)}_{k,abij}$ are further functional derivatives of $\Gamma_k$. Their expressions in the  flat  configuration  ${\bf r}_{k,f}$, Eq.(\ref{rmin}),  are given in Appendix  \ref{Vertexfunctions}, Eqs.({\ref{gamma3}) and ({\ref{gamma4}).

 \subsection{RG equations}
  
For the search of  fixed points   it is convenient to write the flow  equations  in terms of  dimensionless coupling  constants. 
We  thus  define   $\zeta_k^2=Z_k^{-1}\, k^{D-2+z}\overline\zeta_k^2$, $\lambda_k=Z_k^2\, k^{-(D-4+z)}\overline\lambda_k$ and  $\mu_k=Z_k^2\, k^{-(D-4+z)}\overline\mu_k$.  Their flows are given by:

\begin{subequations}
\begin{alignat}{3}
\partial_{t}\overline\zeta_k^2 &= \displaystyle  -(D-2+z+\eta_k)\,\overline\zeta_k^2+ \frac{4\, A_{D}}{D}\times  \nonumber \\
\displaystyle & \hspace{-0.3cm} \Big\{(D-1)\ \dfrac{4\overline \mu_k+\overline \lambda_k D}{2\overline\mu_k+\overline\lambda_kD}\, \overline l_{010}^{D+2} + \nonumber  \\
&  \hspace{-0.3cm} \displaystyle  \frac{6\overline \mu_k+(D+2)\overline  \lambda_k}{2\overline\mu_k+\overline \lambda_k D} \overline l_{001}^{D+2}+(d-D)\overline l_{100}^{D+2}\Big\} \label{eqrg1}   \\
\partial_{t}\overline \mu_k  & =  \displaystyle  (D-4+2\eta_k+z)\overline \mu_k+   \frac{2\, A_{D}}{D(D+2)} \times \nonumber  \\
&   \hspace{-0.3cm} \displaystyle \Big\{4(3\overline \mu_k+\overline \lambda_k)^{2}\, \overline l_{002}^{D+4}+ 4D\, \overline \mu_k (2\overline \mu_k+\overline\lambda_k)\, \overline l_{011}^{D+4} +  \hspace{-0.3cm} \nonumber \\ 
&   \hspace{-0.3cm} 2 \overline \mu_k^{2}(D^{2}+2D-8)\, \overline l_{020}^{D+4} +4\overline \mu_k^{2}(d-D)\, \overline l_{200}^{D+4}\Big\} \label{eqrg2}   \\
\partial_{t}\overline \lambda_k & = \displaystyle  (D-4+z+2\eta_k)\overline \lambda_k+  \frac{2\, A_{D}}{D(D+2)}\times  \nonumber  \\
&   \hspace{-0.3cm} \displaystyle \Big\{-8\overline \mu_k(2\overline \mu_k+\overline \lambda_k)\, \overline l_{011}^{D+4}+ \nonumber \\
&   \hspace{-0.2cm} (d-D)\Big[4\overline \mu_k^{2}+4(D+2)\overline \mu_k\overline \lambda_k+D(D+2)\overline \lambda_k^{2}\Big]\overline l_{200}^{D+4}+ \nonumber \\
&   \hspace{-0.2cm} \Big[4(3D+2)\overline \mu_k^{2}+(D^{2}+D-2)(8\overline \mu_k\overline \lambda_k +D\overline \lambda_k^2)\Big]\overline l_{020}^{D+4}+ \nonumber \\
&   \hspace{-0.2cm} \Big[36\overline\mu_k^{2}+12(D+4)\overline \mu_k\overline \lambda_k+(D^{2}+6D+12)\overline \lambda_k^{2}\Big] \overline l_{002}^{D+4}\Big\} \label{eqrg3} 
\end{alignat}
\end{subequations}
where $A_D^{-1}=2^{D+1}\pi^{D/2}\,\Gamma(D/2)$ and  $\Gamma(\dots)$ is the Euler's gamma function. The equation for $\eta_k$ is too long to be given here while $\tilde\eta_k$  is  discussed   below.  The set  of  equations (\ref{eqrg1})-(\ref{eqrg3}), together with $\eta_k$ and $\tilde\eta_k$,  generalize to the quantum case those  derived in \cite{kownacki09} in the classical case.  \\

\subsection{Threshold functions}
\label{thresholdsection}

 In Eqs.(\ref{eqrg1})-(\ref{eqrg3}) $\overline l_{abc}^{D}$ stands for the dimensionless counterpart of the so-called dimensionful "threshold functions" $l_{abc}^{D}$ -- see Appendix \ref{threshold}:
\begin{equation}
\hskip -0.25cm
\begin{array}{ll}
\displaystyle l^D_{abc}= -\frac{T}{4\, A_{D}}\,\widehat{\partial}_t \sum_{\omega_n}\int_{\bf q}  {{1}\over \left[P_0(\tilde{\bf q})\right]^a \left[P_1(\tilde{\bf q})\right]^b \left[P_2(\tilde{\bf q})\right]^c}
\label{thresholdL}
\end{array} 
\end{equation} 
where  $P_i(\tilde{\bf q})=P(\tilde{\bf q})+m_{ik}^2\, q^{2},  i=0,1,2$ and $P(\tilde{\bf q})=Z_k\, q^{4}+Z_k^{\tau} \omega_n^2+ R_{k}(\tilde{\bf q})$. The  squared masses  $m_{ik}^2$, $i=1,2$  given by    $m_{1k}^2= \mu_k \zeta_k^2$ and  $m_{2k}^2=(2 \mu_k+\lambda_k)\zeta_k^2$ are associated  with  the transverse and longitudinal  phonon  modes while   the squared  mass $m_{0k}^2\equiv 0$ is associated with  the  flexural mode.   The  threshold  functions  are the fundamental ingredient of  the effective average action formalism. First, they control  the integration over   fluctuations  of lower and lower momenta when  $k$  is  decreased.  Second, they encompass the nonperturbative  content of the RG  flow since they  include the propagators  that are  nonpolynomial functions of the masses and thus of the coupling constants.    Third, they  govern the phenomenon of   {\it decoupling}  of massive modes. This  refers to  a situation in which coexist  particles  -- or  excitations  -- with well separated masses, {\it e.g.},  a large one $M$ and a smaller one $m$ the latter   being possibly equal to  zero.   It states  that when the theory is probed at a momentum scale  $p\ll M$   the Green functions  of   the full theory  are identical, up to inverse powers of  $M^2$,  to those  computed from an effective theory where the heavy mass  has been cancelled out.   Within   the effective average action  framework    this  phenomenon occurs  along the RG flow  when the running momentum scale $k$ gets  lower than the large  running mass $M_k$.  Indeed  in this case,  the threshold functions decrease   typically as   inverse  powers  of   $M_k$, see below.        Roughly this   relies on the   fact that  $\partial_k R_k({\bf \tilde q})$ acts  as an ultra-violet cut-off   that  restricts  the values of  $q$ contributing  to the  momenta integrals to the range $[0,k]$; in particular the high momentum  lattice scale $\Lambda$  disappears from the flow.  This   elects  the running scale $k$ as the only reference scale    to which $M_k$ can be compared,  what leads to the announced  result.


\subsection{The temporal anomalous dimension $\tilde{\eta}$}

  We discuss  here the RG flow  of  the field renormalization $Z_k^{\tau}$   and derive the expression of the  running  "temporal" anomalous dimension $\tilde{\eta}_k$  defined by: $\tilde{\eta}_k=-1/Z_k^\tau {\partial_t Z_k^\tau}$.  The quantity  $Z_k^\tau$  is defined  as the coefficient  of   $\omega_m^2$ in the expansion of $\Gamma^{(2)}_k$  in powers of   the external momentum   $p$ and frequency   $\omega_{m}$, at vanishing $\tilde{\bf p}$:
\begin{equation}
Z^{\tau}_k=\displaystyle \lim_{\tilde{\bf p}\to 0}   \frac{d}{d\omega_m^2}\left[\Gamma^{(2)}_{k,D+1,D+1}[{\bf r},{\bf \tilde p},-{\bf \tilde p}]\big |_{{\bf r}_k^f}\right]\  . 
\end{equation}
As a consequence of the discussion  of  subsection \ref{Generalprinciple}  its  flow  is essentially given by the  $\omega_m^2$-derivative of  $\partial_t\Gamma^{(2)}_{k,ij}[{\bf r};{\bf\tilde{p}},-{\bf\tilde{p}}]$, see  Eq.(\ref{gamma2}), taken in the configuration ${\bf r}_k^f$, Eq.(\ref{rmin}).   Since, at this order, the vertices are frequency independent  -- see Appendix  (\ref{Vertexfunctions}), Eqs.(\ref{gamma3})-(\ref{gamma4}) --  the only  contribution to the renormalization of $Z_k^\tau$  can come from the  term  $G\Gamma^{(3)} G\Gamma^{(3)}$ in the right-hand size of  Eq.(\ref{gamma2}),  involving   $G_{k,cd}[{\bf r},\tilde{\bf q}-\tilde{\bf p}]$ that carries the external frequency $\omega_m$. However   the  vertex functions $\Gamma^{(3)}_{k,bjd}$  carry at least one  power of the external momentum ${\bf p}$  so that  the  putative  contribution to the renormalization of  $Z_k^\tau$ vanishes in the limit $\tilde {\bf p}\rightarrow {\bf 0}$ and one has: 
\begin{equation}
\tilde{\eta}_k=0. 
\end{equation}


 \subsection{The  RG flow in the flat phase}

   We  now derive the RG equations  in the flat phase.   They   can be  obtained  by following explicitly  the RG   flow Eqs.(\ref{eqrg1})-(\ref{eqrg3}) in the infra-red,  {\it i.e.} $k\to 0$, regime with initial conditions  taken in this phase. This shows that the  extension  parameter  $\zeta_k$  reaches  a finite value while its  dimensionless counterpart $\overline \zeta_k$ diverges.  On  the other hand,  the coupling constants  $\lambda_k$ and $\mu_k$ vanish while   $\overline\lambda_k$ and $\overline\mu_k$ reach fixed point values. As a consequence the dimensionless masses $\overline m_{1k}^2=\overline\mu_k\overline \zeta_k^2$ and $\overline m_{2k}^2=(2\overline\mu_k+ \overline\lambda_k)\overline \zeta_k^2$ get large in  the infra-red.  We thus meet  the conditions of the decoupling phenomenon where  the phonon modes   get out   the RG flow    leaving  the flexural mode  to govern it.  This provides   a direct way to get  the RG equations in  the flat phase, {\it i.e.}    by considering the limit:     $\overline m_{1k}$, $\overline m_{2k} \to  \infty$.  As this  can be explicitly checked    in the case of the $\Theta$  cut-off -- see Eqs.(\ref{ldedim}) and (\ref{ndedim})   in Appendix (\ref{thresholdtheta})  -- the "massive"  dimensionless threshold functions $\overline{l}^D_{0bc}$ (and $\overline{n}^D_{0bc}$ -- see below) decay as powers of  $\overline m_{1k}^2$  and $\overline m_{2k}^2$, what reflects,  on the dimensionless threshold functions,  the decoupling phenomenon.    In this limit one gets  simplified RG equations   where only  the  part of the flow associated with the flexural mode appears,  what  implies the presence  of  threshold functions $\overline l^D_{a00}$ (and $\overline n^D_{a00}$,  see below)  alone.   The RG flow  for $\zeta_k^2$ is given by:
 \begin{equation}
\partial_{t}\overline\zeta_k^2 = \displaystyle  -(D-2+z+\eta_k)\,\overline\zeta_k^2+ \frac{4\, A_{D}}{D}  (d-D)\overline l_{100}^{D+2}\label{flot0} \ . 
 \end{equation}
This equation  controls   the different phases, and phase transitions, occurring in the phase diagram of membranes.  It admits a  non-trivial  unstable  fixed point $\overline\zeta_*^2$   that corresponds to  the crumpling-to-flat transition  by which we are not interested in here. For $\overline\zeta_k^2>\overline\zeta_*^2$ the flow drives $\overline\zeta_k^2$  toward  its infinite value limit that defines  the flat phase.  The other  RG equations  in this phase  are given by: 
   
\begin{subequations}
\begin{align}
 &\hspace{-0.3cm} \partial_{t}\overline \mu_k = \displaystyle (D-4+2\eta_k+z)\overline \mu_k+\frac{8\, A_{D}}{D(D+2)} \overline \mu_k^{2}(d-D)\, \overline l_{200}^{D+4} \label{flot1}\\
 &\hspace{-0.3cm} \partial_{t}\overline \lambda_k =\displaystyle (D-4+z+2\eta_k)\overline \lambda_k+\frac{2\, A_{D}}{D(D+2)} (d-D)\times \nonumber  \\
 &\hspace{-0.3cm} \displaystyle \hspace{1cm} \Big(4 \overline \mu_k^{2}+4(D+2)\overline \mu_k\overline \lambda_k+D(D+2)\overline \lambda_k^{2}\Big)\overline l_{200}^{D+4}  \label{flot2} \ .
\end{align}
\end{subequations}	
				
As for the  anomalous dimension $\eta_k$ it is given  by : 
\begin{align}
 \displaystyle  \eta_k & =  \displaystyle {2A_{D}\over D(D+2)(2 \overline \mu_k+\overline \lambda_k)}\Big\{2(D+2)(2 \overline \mu_k+\overline \lambda_k)^2\overline n_{200}^{D}+  \nonumber  \\
&\hspace{-0.2cm}  \Big(4(D^2-2D-2)\overline \mu_k(\overline \mu_k+\overline \lambda_k)-D(D+2)\overline \lambda_k^2\Big)\overline l_{100}^{D}\Big\} \ . 
\label{floteta}
\end{align}
 In this last  equation   we have introduced the threshold functions  $\overline n_{abc}^D$  which  are  the dimensionless counterparts  of the dimensionful threshold functions  $n_{abc}^D$ -- see Appendix \ref{threshold}:
\begin{equation}
\displaystyle \hspace{-0.2cm}  n^D_{abc}  = \hspace{-0cm}- \frac{T}{4\, A_{D}} {\widehat{\partial}_t}\sum_{\omega_n}\int_{\bf q}  \hspace{-0cm} {q^2\,   \displaystyle {\partial_{q^2}P(\tilde{\bf q})}\over {\left[P_0(\tilde{\bf q})\right]^a \left[P_1(\tilde{\bf q})\right]^b \left[P_2(\tilde{\bf q})\right]^c}}\ . 
\label{thresholdN}
\end{equation}	

\subsection{Quantum to classical crossover}

The RG equations  Eqs.(\ref{flot1})-(\ref{flot2}) and (\ref{floteta}) account for the effects   of  both classical and quantum  fluctuations when the running scale $k$  varies. In order to quantify the importance of quantum fluctuations, it is convenient to define a dimensionless  effective,  running,   "temperature"  $\overline{T}_k$ through:  
\begin{equation}
T=\Delta_k k^2\,\overline{T}_k 
\label{temp}
\end{equation}
 such that $\overline\beta_k=1/\overline{T}_k$ appears as the dimensionless  thickness of the  time integral in Eq.(\ref{action2}) which characterizes the importance of quantum fluctuations.   The definition of $\overline{T}_k$  leads  to  consider  a typical  momentum -- the thermal momentum scale -- $k_T$   via the relation $\overline{T}_{k_T}\sim 1$ that,  using Eq.(\ref{temp})  and  $\Delta_{k_T}\sim \sqrt{\kappa/\rho}$  yields:
\begin{equation}
k_T\sim \sqrt{\frac{k_BT}{\hbar}\sqrt{\frac{\rho}{\kappa}}}\ . 
\label{thermal}
\end{equation} 
  For  effective  "temperatures" $\overline{T}_k\ll 1$  -- or equivalently for momenta   $k\gg k_T$ -- the  dimensionless thickness   $\overline\beta_k$  of the time integral   extends  from 0 to $\infty$ and quantum  effects are important.  Conversely  for temperatures  $\overline{T}_k\gg 1$ -- or equivalently for  momenta $k\ll k_T$ --   this  thickness $\overline\beta_k$  vanishes and fluctuations are fully classical.   More precisely,    let's  consider  the   action $S \left[{\bf R}\right]$, Eq.(\ref{action}), expressed in terms of  dimensionless quantities: 
  \begin{align}
\displaystyle  S \left[\overline{\bf R}\right]&=\displaystyle \int_0^{\overline\beta_k}  d\bar \tau\:\int d^D\bar x\: \bigg\{{\frac{\bar\rho}{2}(\partial_{\bar\tau} \overline{\bf R})^2}+\frac{\bar \kappa}{2}(\partial_{\bar\gamma}\partial_{\bar\gamma}\overline{\bf R})^2  
 \nonumber  \\
&\hspace{-0.5cm}  \displaystyle+ {\overline{\tilde{\mu\over 4}}}\Big(\partial_{\bar\gamma}\overline{\bf R}.\partial_{\bar\nu}\overline{\bf R}-{\overline{\tilde\zeta}}^{\,2}\delta_{\gamma,\nu}\Big)^2+{\overline{\tilde {\lambda\over 8}}}\Big(\partial_{\bar\gamma}\overline{\bf R}.\partial_{\bar\gamma}\overline{\bf R}-D{\overline{\tilde\zeta}}^{\,2}\Big)^2\bigg\}  \ .
\label{actionbis}
\end{align}
 When  $\overline{T}_k\gg 1$ or, equivalently,   when  $\overline\beta_k$ vanishes   the  only configurations of $\bf R(\bar{\bf x},\bar\tau)$ that contribute to the functional integral Eq.(\ref{fonctionnal2})   are  those  that minimize  locally the kinetic term  $(\partial_{\bar\tau} {\overline{\bf  R}})^2 $, {\it i.e.} those that are time-independent. As a consequence  the effective average action $\Gamma_k\left[\bar{\bf r}\right]$  resulting from this functional integral   is also time-independent.   The    ${\overline\beta_k}$   term     then factors out and  the  effective   action takes the form:
\begin{align}
\displaystyle\Gamma_k\left[\bar{\bf r}\right]=&\displaystyle  {\overline\beta_k}   \int d^D\overline x\:\bigg\{  \frac{Z_k}{2}(\partial_{\bar\gamma}\partial_{\bar\gamma}{\bar{\bf r}})^2\nonumber  \\
&\hspace{-0.8cm}+ {\overline\mu_k \over 4}\big(\partial_{\bar\gamma}{\bar{\bf r}}.\partial_{\bar\nu}{\bar{\bf r}}-\overline\zeta_k^{\, 2}\delta_{\gamma,\nu}\big)^2 +{\overline \lambda_k \over 8}\big(\partial_{\bar\gamma}{\bar{\bf r}}.\partial_{\bar\gamma}{\bar{\bf r}}-D \overline\zeta_{k}^{\, 2}\big)^{2}\bigg\}\ . 
\label{action4}
\end{align}

Redefining  $\bar{\bf r} \to \overline\beta_k^{-1/2}\,\bar{\bf r}$ and  introducing  the  "classical" coupling constants $\overline\lambda_{k}^{cl}=\overline\beta_k^{-1} \lambda_{k}$, $\overline\mu_{k}^{cl}=\overline\beta_k^{-1} \mu_{k}$ and $\overline\zeta_{k}^{cl}=\overline\beta_k^{1/2} \zeta_{k}$, one gets the effective action of classical  membranes:
\begin{align}
\displaystyle\Gamma_k\left[\bar{\bf r}\right]=&\displaystyle \int d^D\overline x\:\bigg\{\frac{Z_k}{2}(\partial_{\bar\gamma}\partial_{\bar\gamma}{\bar{\bf r}})^2\nonumber  \\
&\hspace{-0.8cm}+ {\overline\mu_{k}^{cl} \over 4}\big(\partial_{\bar\gamma}{\bar{\bf r}}.\partial_{\bar\nu}{\bar{\bf r}}-{\overline\zeta_{k}^{cl}}^{2}\delta_{\gamma,\nu}\big)^2 +{\overline\lambda_{k}^{cl} \over 8}\big(\partial_{\bar\gamma}{\bar{\bf r}}.\partial_{\bar\gamma}{\bar{\bf r}}-D{\overline\zeta_{k}^{cl}}^{2}\big)^{2}\bigg\}\ . 
\label{action5}
\end{align}
This illustrates  the well-know phenomenon of  thermal  dimensional reduction from a $(D+1)$-dimensional to a $D$-dimensional model  associated with the quantum to classical crossover that occurs when   the  temperature increases \cite{chakravarty91a}.  

It is instructive  to interpret this phenomenon from the RG point of view.  As said in section (\ref{cutoffsection}),  we have chosen a cut-off that acts both on momenta and frequencies.  Using the  $\Theta$ cut-off  consists in taking, in Eq.(\ref{cutoff2}), ${\cal R}(Y)=(Y^{-1}-1)\Theta(1-Y)$ which leads to: 
\begin{equation}
R_k(\tilde{\bf  q})=Z_k \left(k^4-\Big(q^4+{\omega_n^2\over  \Delta_k^2}\Big)\right) \Theta\left(1-\Big({q^4\over k^4}+{\omega_n^2\over  \Delta_k^2k^4}\Big)\right) \, .  
\label{cutoffexplicit}
\end{equation}

With this form of cut-off   it is easy to verify that the term  $\partial_t R_k({\tilde{\bf q}})$  entering in the threshold functions  $l_{abc}^D$ and $n_{abc}^D$ enforces strictly \footnote{Another form of cut-off would impose the same kind of constraints but  softened.  }  that  only the range of momenta/frequencies  $0\le q^4/k^4+\omega_n^2/(\Delta_k^2 k^4) \le 1$ contributes   to the RG flow.  Once the integral is  performed on momentum $q$  the constraint on $\omega_n$ reads: $\omega_n/(\Delta_k k^2)\le 1$.  Using   $\omega_n=2\pi n T$ and the relation Eq.(\ref{temp})   this inequality  implies that  the frequency  modes that effectively contribute to the RG flow are those  characterized by  integers  $n$ obeying: $n\le n_{max}= \lfloor1/(2\pi\overline{T}_k)\rfloor$ where $\lfloor\dots\rfloor$ denotes the floor function. Finally  the RG flow of  $\overline{T}_k$ is easily obtained from Eq.(\ref{temp}) and the flow of $\Delta_k$:
\begin{equation}
\partial_t \overline{T}_k= - z\   \overline{T}_k 
\end{equation} 
which  means that  $\overline{T}_k$ is, as expected,  a relevant variable \footnote{We recall that the infra-red behaviour corresponds to $k\to 0$ and thus $t\to-\infty$.}.  Thus, as   $k$ decreases,   the high-frequency modes   are progressively integrated out and disappear   from the RG flow.    This can be seen, again,  as a kind of decoupling of the  non-vanishing (or "massive")  Matsubara frequencies  at low momenta  leading to a situation dominated by the classical  -- $\omega_n=0$ --   mode.

 \subsection{The RG  flow at  vanishing  temperature}
		
 We   now discuss the RG equations in the flat phase.  We start by considering the  limit of  vanishing effective  temperatures   $\overline{T}_k\ll  1$ or   large  running momentum scale    $k\gg k_T$.  In this case all frequency  modes  contribute. Technically   the sums  over frequencies in  the   dimensionless threshold functions $\overline l^D_{abc}$   and $\overline n^D_{abc}$, see  Appendix \ref{Dimthresfunct}, Eqs.(\ref{ldedim}}) and (\ref{ndedim}),    are   replaced by integrals  over continuous frequencies:
 \begin{equation}
 \overline{T}_k\sum_{\overline\omega_n}\to \int_{-\infty}^{\infty} {d\overline{\omega}\over 2\pi}\ .
 \end{equation}
 
Replacing, in  the RG flow equations, Eqs.(\ref{flot1})-(\ref{flot2}) and (\ref{floteta}),   the  threshold functions   by their simplified expressions, Eqs.(\ref{ldplate0})-(\ref{ndplate0}),  and   $z$ by its value $z=2-\eta_k/2$  one obtains   the equations for any dimension $D$. For simplicity we concentrate here on the physical  $D=2$  case in which one has: 
\begin{subequations}
\begin{alignat}{3}
 \partial_t \overline \mu_k  & =  \displaystyle  {3\over 2}\eta_k \overline \mu_k+  {1\over 32 \pi} (d-2){\overline\mu_k^2} \Big(1-{\eta_k\over 8}\Big) \label{zeroT1}   \\
\nonumber   \\
\partial_t \overline \lambda_k &= \displaystyle  {3\over 2}\eta_k \overline \lambda_k \nonumber \\
 &  \displaystyle + {1\over 32 \pi}(d-2)(2\overline\lambda_k^2+4\overline\lambda_k\overline\mu_k+\overline\mu_k^2)\  \Big(1-{\eta_k\over 8}\Big)    \label{zeroT2}   \\
 \nonumber   \\
\hspace{0.3cm}\eta_k &= \displaystyle{48 \overline \mu_k(\overline\lambda_k+\overline\mu_k)\over   9\overline \mu_k(\overline\lambda_k+\overline\mu_k)+128 \pi (\overline\lambda_k+2\overline\mu_k)}
 \label{zeroT3} \ . 
\end{alignat}
\end{subequations}

These equations are the nonperturbative generalization of  those  found previously  perturbatively \cite{sanjose11,kats14,guinea14}.

\subsubsection{Large $d$ expansion} As argued in the general discussion of section (\ref{properties})  our  equations  should reproduce the leading order in a $1/d$ expansion. In this case   one   assumes   that  the  coupling constants  behave as   $1/d$.  At leading order in $1/d$  Eqs.(\ref{zeroT1})-(\ref{zeroT3}) then simplify  into, after the rescaling   $(\overline \mu_k,\overline \lambda_k)\to (\overline \mu_k/d,\overline \lambda_k/d)$:
\begin{subequations}
\begin{align}
\hspace{0.5cm}  \partial_t \overline \mu_k &=  \displaystyle   {1\over 32 \pi}{\overline\mu_k^2} \hspace{4cm} \label{zeroTlarged1}\\
 \nonumber \\
 \partial_t \overline \lambda_k& =   \displaystyle  {1\over 32 \pi}(2\overline\lambda_k^2+4\overline\lambda_k\overline\mu_k+\overline\mu_k^2)   \label{zeroTlarged2}\ \\
\nonumber \\
\hspace{0.3cm}\eta_k& = \displaystyle {3\over 8\pi d}  {\overline \mu_k(\overline\lambda_k+\overline\mu_k)\over (\overline\lambda_k+2\overline\mu_k)}
\label{zeroTlarged3}\ \ . 
\end{align}
\end{subequations}
It is convenient  to write the RG equations (\ref{zeroTlarged1})-(\ref{zeroTlarged2}) in terms of  the Young modulus $\overline{\mathcal{Y}}_k=4\overline\mu_k(\overline\mu_k+\overline\lambda_k)/(2\overline\mu_k+\overline\lambda_k)$. One obtains:
\begin{equation}
\partial_t \overline {\mathcal{Y}}_k =  \displaystyle {3\over 128 \pi}  \overline{\mathcal{Y}}_k^2\\
\label{Ylarged}
\end{equation}
which,  restoring  the coupling constants  $\kappa$ and $\rho$  coincides with the expression obtained by Guinea \etal \cite{guinea14}  and  -- up to a factor 2 already noted in  \cite{guinea14} -- with the equation derived by San Jose \etal \cite{sanjose11}.  Note that   $\eta_k$  provides   the flow of  the bending rigidity modulus $\kappa$ in  the original action Eq.(\ref{action})  through  the relation $\eta_k=-\kappa^{-1}\partial_t \kappa$.  One gets \footnote{In the following  equation the rescaling has not been  made.}:
\begin{equation}
\partial_t   \kappa = - \displaystyle {3\over 32 \pi}\,   \overline{\mathcal{Y}}_k\, \kappa\\
\label{Ylarged}
\end{equation}
which,   restoring  the coupling constants  $\kappa$ and $\rho$,    coincides again,  up to a factor 2,   with the equation  obtained in \cite{sanjose11}.

\subsubsection{ Weak-coupling expansion}

The  equations Eqs.(\ref{zeroT1})-(\ref{zeroT3}) can also be  expanded  in powers of the coupling constants $\overline\lambda_k$ and $\overline\mu_k$.  One finds, at leading  non-trivial order: 
\begin{subequations}
\begin{align}
\partial_t \overline \mu_k &=  \displaystyle  {3\over 2}\eta_k \overline \mu_k+  {1\over 32 \pi} (d-2){\overline \mu_k^2} \label{zeroTweak1} \\
\nonumber  \\
\partial_t \overline \lambda_k& = \displaystyle  {3\over 2}\eta_k \overline \lambda_k+  {1\over 32 \pi}(d-2)(2\overline\lambda_k^2+4\overline\lambda_k\overline\mu_k+\overline\mu_k^2) \label{zeroTweak2} \\
\nonumber  \\
\eta_k& = \displaystyle {3\over 8\pi}{\overline \mu_k(\overline \mu_k+\overline \lambda_k)\over \overline \lambda_k+2 \overline\mu_k }\ . \label{zeroTweak3}
\end{align}
\end{subequations}
 In the particular case where $d=3$, and  restoring  the coupling constants  $\kappa$ and $\rho$, these equations  coincide with those  derived by Kats and Lebedev in  \cite{kats14}. There is, however,  a drastic  difference that concerns the sign of  $\eta_k$ that we find to be  reversed  with respect to theirs.  This leads to a major difference as for the behaviour of  membranes at long distances -- see below.

\subsubsection{RG flow in $D=2$ and $d=3$} 

  We  have  studied  the RG equations   Eqs.(\ref{zeroT1})-(\ref{zeroT3}),  {\it i.e.}  searched  for  the fixed points and analyzed  the structure of the flow, in  the physical $D=2$ and $d=3$ case.  They  display one  physical fixed  point  which  corresponds to the gaussian one  with $\overline \mu=\overline \lambda=0$.    This fixed point is {\it attractive}   in the infra-red, see Fig.(\ref{Figflow}),   in agreement with the fact that  $D=2$  is,   according to previous considerations,   supposed to be the upper critical dimension of  the theory.  This contradicts   the result   obtained in \cite{kats14} that   the gaussian fixed point is repulsive.

\begin{figure}[h]
\includegraphics[scale=1]{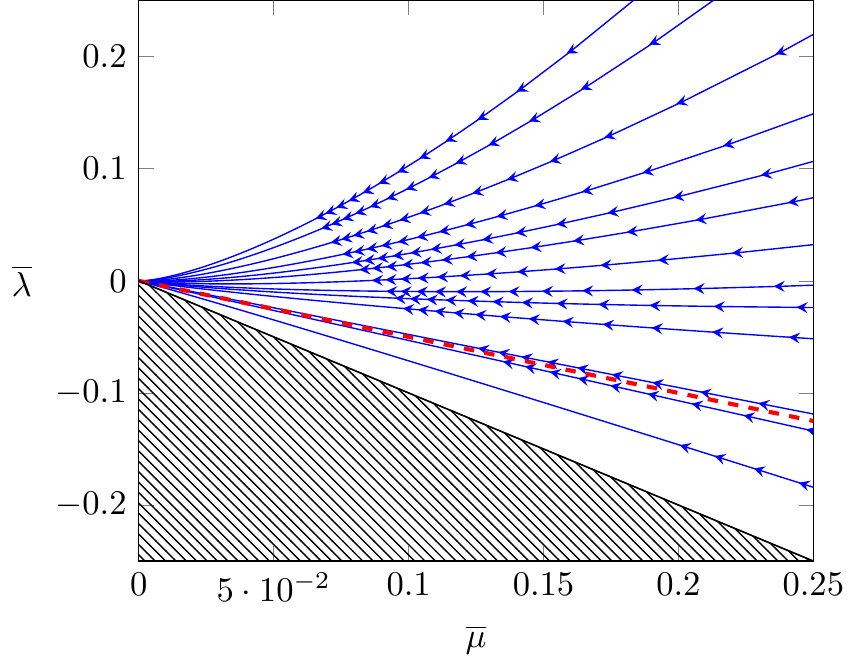}
\caption{The RG flow in the $(\overline\lambda,\overline\mu)$ plane. The shaded area corresponds to the forbidden region $\overline\lambda+\overline\mu<0$; the bordeline $\overline\lambda=-\overline\mu$ is repulsive as the line $\overline\mu=0$. The dashed line corresponds to  the attractive line $\overline\lambda=-\overline\mu/2$ along which all trajectories converge. }
\label{Figflow}
\end{figure}


Now since, in the infra-red,   the RG flow is essentially controlled by the gaussian fixed point we can restrict  our study  to the vicinity of this fixed point and consider the perturbative limit  Eqs.(\ref{zeroTweak1})-(\ref{zeroTweak3}) of our equations.  They  read in $d=3$:
\begin{subequations}
\begin{align}
\partial_t \overline \mu_k &=\displaystyle   {\overline \mu_k^2 (19 \overline\lambda_k+20 \overline\mu_k)\over 32 \pi  (\overline\lambda_k+2 \overline\mu_k)} \label{zerot2D3d1} \\
\nonumber \\
\partial_t \overline \lambda_k& = \displaystyle    {2\overline\lambda_k^3+26\overline\lambda_k^2\overline\mu_k+27 \overline\lambda_k\overline\mu_k^2+2\overline\mu_k^3\over   32 \pi (\overline\lambda_k+2 \overline\mu_k)}   \label{zerot2D3d2} \\
\nonumber \\
  \eta_k& = \displaystyle {3\over 8\pi}{\overline \mu_k(\overline \mu_k+\overline \lambda_k)\over \overline \lambda_k+2 \overline\mu_k }  \label{zerot2D3d3} \ . 
\end{align}
\end{subequations}

 It is instructive to compute  the  flow of  the ratio $\overline\mu_k/\overline\lambda_k$ which given by:  
 \begin{equation}
\partial_t\Big({\overline\mu_k\over \overline\lambda_k}\Big)=-{\overline\mu_k\over 32 \pi \overline\lambda_k^2}  ({\overline\lambda_k+\overline\mu_k})(2{\overline\lambda_k+\overline\mu_k})\ . 
\end{equation}
 It is then easy to shows that  the RG flow admits one attractive  line, $\overline\lambda=-\overline\mu/2$, see Fig.(\ref{Figflow}), and  two repulsive ones, $\overline\lambda=- \overline\mu$  and  $\overline\mu=0$.      The  behaviour of the flow  in the vicinity  of the gaussian  fixed point  along the attractive line is given by:
\begin{subequations}
\begin{align}\partial_t \overline \mu_k&=\displaystyle   {7\over 32 \pi}  \overline \mu_k^2 \label{flotpert1} \\
\nonumber \\
\eta_k &= \displaystyle {\overline \mu_k\over 8\pi}  \label{flotpert2}\ .  
\end{align}
\end{subequations}

The effective coupling constant $\overline \mu_k$  decreases  at long distances,  again   in agreement with the fact that  $D=2$ is expected to be the  upper critical dimension of  quantum membranes and  in disagreement with \cite{kats14} where  an increasing effective coupling constant is found  at long distances.    Integrating then  the flow equations  Eqs.(\ref{flotpert1})-(\ref{flotpert2})  provides the  expressions  of the momentum dependent  bending rigidity  $\kappa(q) \sim \vert \ln q\vert^{4/7}$,  in agreement with \cite{gonzalez14},  and the momentum dependent  coupling $\overline \mu(q)\sim \vert \ln q\vert^{-1}$.    These logarithmic behaviours, together with the considerations of section (\ref{spectrum}) about the fluctuations in the flat phase,   imply that polymerized membranes  are not destabilized by quantum fluctuations at $T=0$.      Note finally that, as for the  momentum dependent anomalous dimension $\eta(q)$,  we also predicts a  vanishing -- logarithmic -- behaviour  $\eta(q)\sim    \vert \ln q\vert^{-1}$  that contrasts with the prediction of \cite{amorim14} that  leads to a finite value -- $\eta =2$ --   at vanishing temperature. 

Again one can rewrite Eqs.(\ref{zerot2D3d1}) and (\ref{zerot2D3d2}) in terms of   the Young modulus: 
 \begin{equation}
\partial_t \overline{\mathcal{Y}}_k =  \displaystyle {21\over 128 \pi}  \overline{\mathcal{Y}}_k^2
\label{Y}
\end{equation}
that    integrates  into:  
\begin{equation}
\overline{\mathcal{Y}}_k =  \displaystyle {\overline{\mathcal{Y}}_{\Lambda}\over 1- \displaystyle {21\,  t\over 128 \pi} \overline{\mathcal{Y}}_{\Lambda}}  \  . 
\label{Yint}
\end{equation}
 The Young modulus is generally    used  to define a  quantum Ginzburg momentum scale $k_G^q$ that separates a strong   from  a weak  coupling regime.  Defining this scale as the one at which    the Young modulus looses half its initial value:
\begin{equation}
{\vert\delta\overline{\mathcal{Y}}\vert\over\overline{\mathcal{Y}}_{\Lambda}}={\vert\overline{\mathcal{Y}}_k-\overline{\mathcal{Y}}_{\Lambda}\vert\over\overline{\mathcal{Y}}_{\Lambda}}\simeq {1\over 2}
\label{criterion}
\end{equation}
one obtains:
\begin{equation}
k_G^q\sim \Lambda\  e^{\displaystyle -{64 {\pi}\over {21\,  \overline{\mathcal{Y}}_{\Lambda}}}}\ . 
\label{qGq}
\end{equation}

Our expression  is close to that  obtained in  the large $d$ approach of  \cite{guinea14}  with a  change  of coefficient in the argument of the exponential \footnote{The coefficient  $64 \pi/21$  here replaces   the coefficient $64 \pi/3$  in \cite{guinea14}.}  due to the non-vanishing anomalous dimension within our computation.

\vspace{0.2cm} 

 \subsection{The RG flow at high  temperatures}
 
\subsubsection{General equations } 

 We now consider  the limit of high effective temperatures    $\overline{T}_k\gg 1$  -- or    low momentum   scales  $k\ll k_T$. As said previously, in this case, due to the thermal decoupling of high frequency modes,   only   the  vanishing frequency mode  contributes  to  the RG flow. Technically the sums  in the threshold functions $\overline l^D_{abc}$   and $\overline n^D_{abc}$, see Eqs.(\ref{flatl}) and (\ref{flatn}),  reduce to the value $\omega_n=0$.  In this case the threshold functions are given by Eqs.(\ref{flathigha})-(\ref{flathighb}).    Using these quantities and the set of  classical coupling constants :  $\overline\lambda_{k}^{cl}=\overline\beta_k^{-1} \overline\lambda_{k}$, $\overline\mu_{k}^{cl}=\overline\beta_k^{-1} \overline\mu_{k}$ and  $ \overline{\cal Y}_k^{cl} =\overline\beta_k^{-1} \overline{\cal Y}_k$   one obtains from Eqs.(\ref{flot1})-(\ref{flot2}) and  (\ref{floteta}): 
\begin{subequations}
\begin{align}
&\partial_t \overline{\mu}_k^{cl}=(D-4+2 \eta_k)\overline{\mu}_k^{cl} +\displaystyle{4\,  (d-D)\  {\overline{\mu}_k^{cl}}^2\ \widetilde A_D} \label{flat1}  \\
\nonumber \\
&\partial_k \overline{\lambda}_k^{cl}=(D-4+2 \eta_k) \overline{\lambda}_k^{cl} + (d-D)\times  \nonumber \\
\nonumber\\
&\displaystyle \ \ {\big(4{\overline{\mu}_k^{cl}}^2+4(D+2) \overline{\mu}_k^{cl} \overline{\lambda}_k^{cl}+D(D+2) {\overline{\lambda}_k^{cl}}^2\big) \widetilde A_D} \label{flat2} \\ 
\nonumber \\ 
&\displaystyle \eta_k={16 (D+4)(D^2-1)\overline{\cal Y}_k^{cl}  A_D \over D^2(D^2+6D+8)+16(D^2-1)\overline{\cal Y}_k^{cl} A_D}\   
\label{flat3}
\end{align}
\end{subequations}
with $\widetilde{A}_D=16 A_D(8+D-\eta_k)/D(D+2)(D+4)(D+8)$. 

These equations coincide   exactly  -- up to redefinitions of the coupling constants -- with those derived  in \cite{kownacki09} for classical membranes.     They have been shown to agree  with those obtained from  a   perturbative expansion performed around $D=4$  \cite{aronovitz88,aronovitz89}  as well as with those  obtained   within a  large $d$ approach \cite{david88,aronovitz89,ledoussal92}.    Thus our formalism smoothly interpolates between the  full quantum situation and the  full classical one.

\subsubsection{RG flow in $D=2$ and $d=3$} 
   
In  physical dimensions  the RG  flow is given by:
\begin{subequations}
\begin{align}
\partial_t \overline{\mu}_k^{cl}&=\displaystyle 2(\eta_k-1)\overline{\mu}_k^{cl}  +  {{\overline{{\mu}}_k^{cl}}^2\over 60 \pi} (10- \eta_k) \label{eqcl1} \\
\nonumber \\
\partial_k \overline{\lambda}_k^{cl}&=2( \eta_k-1) \overline{\lambda}_k^{cl}+ 
\nonumber\\ 
& \displaystyle  {1\over 60 \pi}\big(2{\overline{\lambda}_k^{cl}}^2+ 4 \overline{\mu}_k^{cl}  \overline{\lambda}_k^{cl}+{\overline{\mu}_k^{cl}}^2\big) (10- \eta_k)  \label{eqcl2}\\
\nonumber\\ 
\eta_k&=\displaystyle {6  \overline{\mu}_k^{cl} ( \overline{\lambda}_k^{cl} + \overline{\mu}_k^{cl})\over \overline{\mu}_k^{cl}( \overline{\lambda}_k^{cl}+\overline{\mu}_k^{cl}) +4\pi ( \overline{\lambda}_k^{cl}+2\overline{\mu}_k^{cl})  }\label{eqcl3}  \  . 
\end{align}
\end{subequations}
 These equations have been solved in \cite{kownacki09}. They admit a stable fixed point  with coordinates  $\overline{\mu}_*^{cl}\simeq 6.21$, $\overline{\lambda}_*^{cl}\simeq -3.10$ with the  associated critical exponent  $\eta^*\simeq 0.849$.

 The flow of the  ratio $\overline{\mu}_k^{cl}/\overline{\lambda}_k^{cl}$ is given by:  
\begin{align}
\displaystyle \partial_t \bigg({\overline{\mu}_k^{cl}\over \overline{\lambda}_k^{cl}}\bigg)=&-\displaystyle { \overline{\mu}_k^{cl}(\overline{\lambda}_k^{cl}+\overline{\mu}_k^{cl})(2\overline{\lambda}_k^{cl}+\overline{\mu}_k^{cl})\over 15 \pi {\overline{\lambda}_k^{cl}}^2}\   \times \nonumber \\
&\displaystyle {\overline{\mu}_k^{cl}(\overline{\lambda}_k^{cl}+\overline{\mu}_k^{cl})+10\pi (\overline{\lambda}_k^{cl}+2\overline{\mu}_k^{cl})\over \overline{\mu}_k^{cl}(\overline{\lambda}_k^{cl}+\overline{\mu}_k^{cl})+4\pi(\overline{\lambda}_k^{cl}+2\overline{\mu}_k^{cl})}
\end{align}
One deduces that the  RG flow admits, as in the vanishing temperature case, one attractive  line, $\overline\lambda=-\overline\mu/2$, and      two   repulsive  ones $\overline\lambda=- \overline\mu$ and  $\overline\mu=0$   that are  identical to those  found in this last case.

 Here,  again, one can exhibit a characteristic -- classical -- Ginzburg momentum scale separating a strong  from  a  weak  coupling regime. We consider  the flow of the dimensionful Young modulus  \footnote{In the quantum case, $D=2$ being the upper critical dimension  there is no distinction between  $\mathcal{Y}_k$ and  $\overline{\mathcal{Y}}_k$.} at dominant  order in $\mathcal{Y}_k^{cl}$.   It  is given by:
\begin{equation}
\partial_t \mathcal{Y}_k^{cl}=\displaystyle{7\over 8\pi} {{\mathcal{Y}_k^{cl}}^2\over k^2}
\label{Y}
\end{equation}
that integrates into:
\begin{equation}
\mathcal{Y}_k^{cl}=\displaystyle {\mathcal{Y}_{\Lambda}^{cl}\over 1-\displaystyle {7\over 8 \pi} \mathcal{Y}^{cl}_{\Lambda} \Big({1\over 2\Lambda^2}-{1\over 2 k^2}\Big)}\ .
\end{equation}
From  Eq.(\ref{criterion}), and  restoring the physical units, we  get:
\begin{equation}
k_G^{cl}=\sqrt{\frac{7 {\mathcal{\widetilde Y}}^{cl}_{\Lambda}}{8\pi}\frac{k_BT}{\kappa^2}}\ .
\end{equation}

\section{Application to  graphene}

We  now illustrate  the different crossovers encountered  in   previous sections  in the context of  the physics of free standing graphene. 

\subsection{Initial conditions}

We first specify the  initial conditions of the RG flow.  We take  as  microscopical characteristics of graphene  (see, e.g.,  \cite{amorim16} and references therein) : $ \kappa\simeq1\,\text{eV}, \rho\simeq7.6\times10^{-7}\text{ kg.m}^{-2},  \mu\simeq3\lambda\simeq9\,\text{eV.\AA}^{-2}$,  the lattice parameter  $a\simeq 1.4\,\text{\AA}$ and $\mathcal{Y}\simeq 20.6\, \text{eV.\AA}^{-2}$. They  are related to the  dimensionless bare quantum coupling constants  at the lattice scale $\Lambda=a^{-1}$: $\overline \lambda_{\Lambda}=\hbar\,  \lambda /(\rho^{1/2} \kappa^{3/2})$ and  $\overline \mu_{\Lambda}=\hbar\,  \mu /(\rho^{1/2} \kappa^{3/2})$.  As the  classical  bare coupling  constants  they are  defined  by    $\overline\lambda_{\Lambda}^{cl}=\overline{T}_{\Lambda} \overline\lambda_{\Lambda}$  and $\overline\mu_{\Lambda}^{cl}=\overline{T}_{\Lambda} \overline\mu_{\Lambda}$. The bare  dimensionless temperature  $\overline{T}_{\Lambda}$  is related to  $\Delta_{\Lambda}=\sqrt{\kappa/\rho}$ by Eq.(\ref{temp}). This leads to: 
\begin{equation}
\overline{T}_\Lambda=\sqrt{\dfrac{\rho}{\kappa}}\frac{k_BT}{\hbar}a^2\ . 
\end{equation}

\subsection{Crossover momenta}

Once  the temperature $T$  has been  chosen the crossover  scales  are completely determined.  One takes  $T=10\,$K  since the temperature below which quantum fluctuations are expected to be dominant is  estimated to be around 70 - 90 K \cite{amorim16}.  For an initial temperature of 10$\,$K   the thermal  momentum  scale  $k_T=a^{-1}e^{-t_T}$  is given by :  
\begin{equation}
k_T= \sqrt{\frac{k_BT}{\hbar}\sqrt{\frac{\rho}{\kappa}}}\sim  a^{-1} e^{-1.44}\sim 0.17\text{\AA}^{-1}
\label{thermal}
\end{equation}    
 that corresponds  to  a  length of a few lattice spacing.

As for the quantum Ginzburg momentum scale it is given by: 
\begin{equation}
k_G^q=  \Lambda\  e^{\displaystyle -{\dfrac{64\pi\kappa^{\frac{3}{2}}\sqrt{\rho}}{21\mathcal{Y}\hbar}}}\sim a^{-1} e^{-15.4}  \ll 1
\end{equation}
 which is  very small  with respect to $k_T$,  in agreement  with  \cite{guinea14}  but in contradiction with    \cite{amorim14}  that predicts  $k_G^q\sim 0.1\, \text{\AA}^{-1}$.  The origin of this contradiction  lies on  the fact that  $D=2$ being, within our computation,  the upper critical dimension of  quantum membranes,  $k_G^q$ takes the  usual -- exponentially decreasing  -- form of a mass-gap in such a  dimension, see Eq.(\ref{qGq}). This is not the case in \cite{amorim14}. 

Since  $k_G^q \ll k_T$, {\it i.e.}  the weak/strong coupling  quantum  crossover occurs far  in the region where thermal fluctuations already  dominate,  it cannot be observed and has no role in  the physical behaviour of the graphene sheet. 

In the classical regime  the  Ginzburg momentum scale  is given by:   
\begin{equation}
k_G^{cl}=\sqrt{\frac{7 {\mathcal{\widetilde Y}}^{cl}_{\Lambda}}{8\pi}\frac{k_BT}{\kappa^2}} \simeq  a^{-1} e^{-2.30}\sim 0.07 \text{\AA}^{-1} \ . 
\end{equation}
which implies  that  $k_G^{cl}<k_T$.

\subsection{Crossover  behaviour of coupling constants}

 The  behaviour of the system under  the RG flow is as  follows:    starting   in a quantum   regime    governed   by  a   gaussian fixed point,  it  goes  through   the quantum to classical crossover.   It   then   enters   a   classical  weak-coupling  regime controlled by  a  gaussian fixed point  and   is   finally  driven toward a strong coupling  infra-red   classical  regime governed by a non-trivial    fixed point.   

These  behaviours    are  first  illustrated  on the RG  flow of the  coupling constants  $\lambda$ and  $\mu$. 
In  Fig.(\ref{qcoup})  we have displayed  the flow of   the {\it quantum}  coupling constants  $\overline\lambda_k$ and  $\overline\mu_k$ as functions of the RG "time"  $-t=\ln\Lambda/k$.  Starting at finite values  they  first decrease  slowly toward    their  (vanishing)   gaussian fixed point  values, their  behaviour being essentially controlled  by the (anomalous) dimensional part of the   vanishing temperature RG flow, see Eqs.(\ref{zeroT1})-(\ref{zeroT3}).  This vanishing  behaviour  is  then amplified when the  quantum to  classical  RG  time $t_T$ is reached since,  there,  the flow starts   to be controlled by  the  finite temperature RG equations Eqs.(\ref{flot1})-(\ref{flot2}) and (\ref{floteta})  and, in particular, by non-trivial  part of the flow proportional to the   effective  temperature $\overline T_k$ carried by the threshold functions,   see Eqs.(\ref{flathigha})-(\ref{flathighb}).

		 
\begin{figure}[!h]
\hskip  -0cm
\includegraphics[scale=0.52]{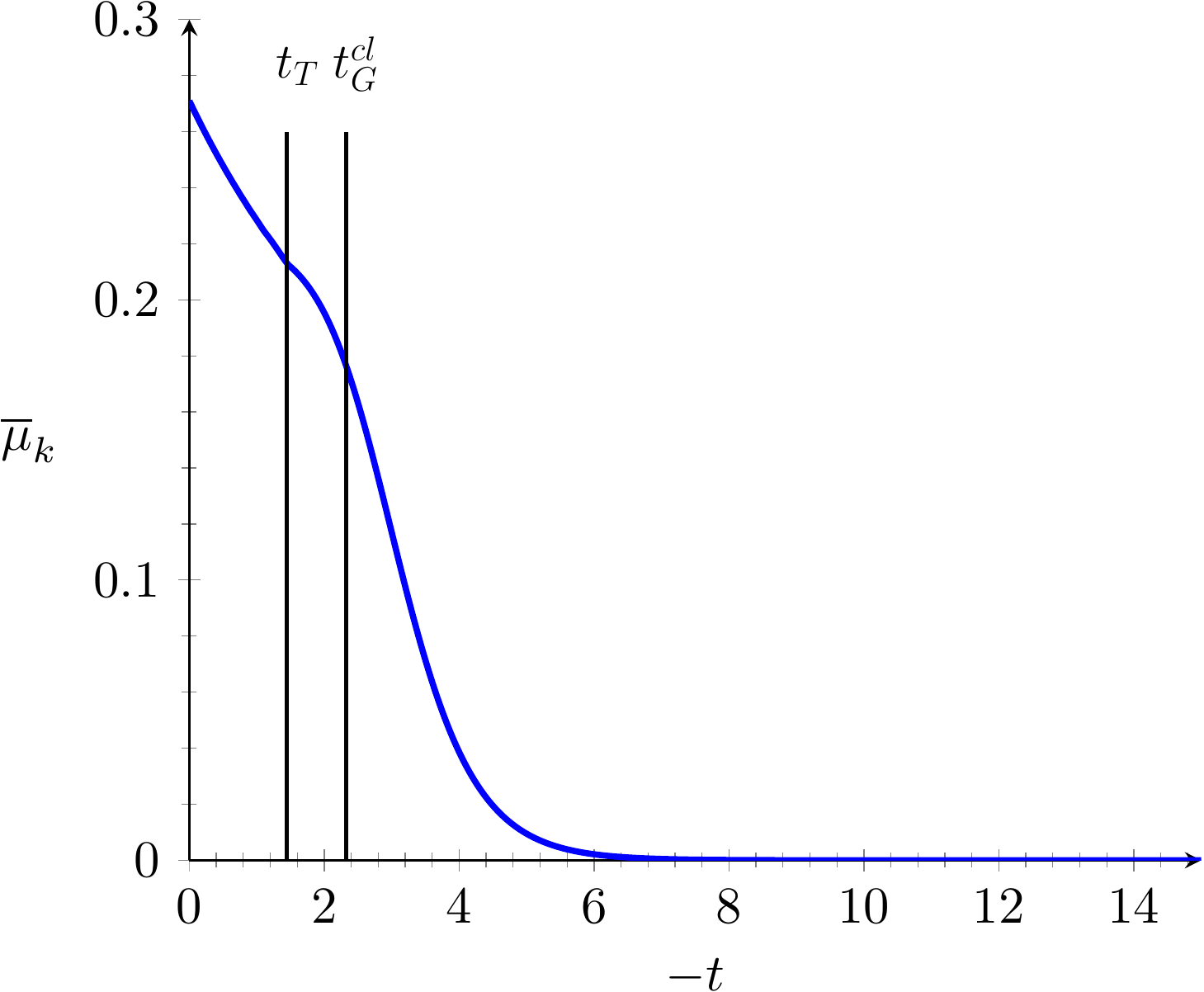}

\hskip  -0.8cm
 \includegraphics[scale=0.55]{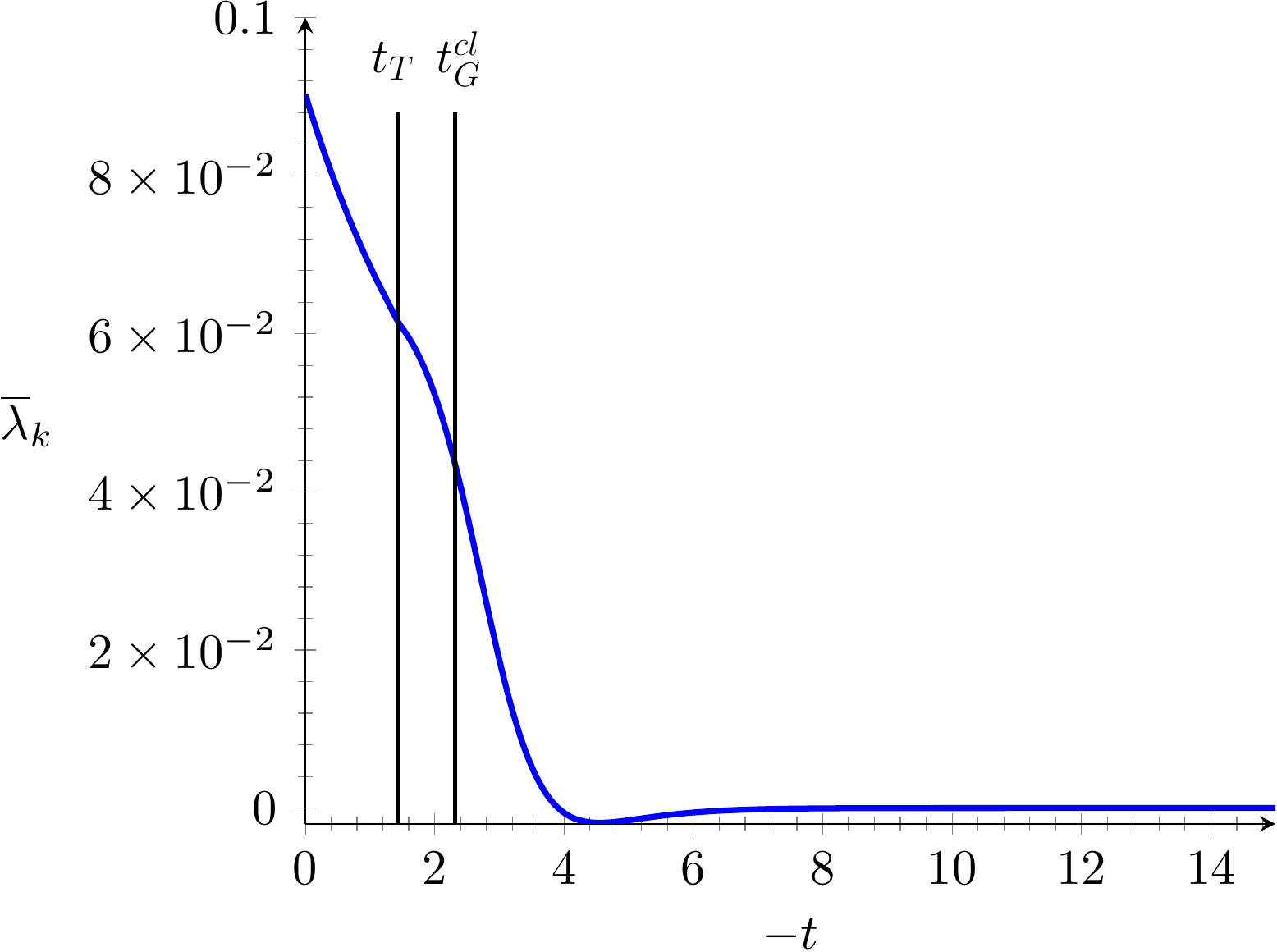}
\caption{Evolution of the dimensionless {\it quantum}  coupling constants $\overline\mu_k$  and $\overline\lambda_k$  with  RG time $-t=\ln \Lambda/k$.   The left vertical   line  corresponds to the quantum to classical   crossover  momentum scale $k_T$  that is located at   $-t_T=1.44$; the right vertical  line corresponds to the classical Ginzburg momentum scale $k_G^{cl}$  that is located at  $-t_G^{cl}=2.30$.  $t_T$ marks a crossover between a regime controlled by the quantum gaussian fixed point and a regime where  the  vanishing behaviour is enhanced by the effective temperature  $\overline T_k$.}
\label{qcoup}
\end{figure}


In  Fig.(\ref{clcoup})  we have  displayed  the flow  of the {\it classical}  dimensionless coupling constants  $\overline\mu_k^{cl}$  and $\overline\lambda_k^{cl}$  as functions of the RG "time"  $-t=\ln\Lambda/k$. These coupling  constants undergo -- roughly -- a symmetrical behaviour.  From their  very  definition  including  thermal factor $\overline\beta_k^{-1}$,    they  start from  almost vanishing values  and    first  increase    slowly   with the  effective  temperature  $\overline T_k$.  Then, when  the  quantum to  classical  RG  time $t_T$ is reached,  their behaviour  begins   to be  controlled by  the classical RG Eqs.(\ref{eqcl1})-(\ref{eqcl2})  and  (\ref{eqcl3})     which   amplifies this  tendency.      At $t_G^{cl}$,  the coupling constants  enter   a   regime where the non-linearities of  the flow start to play a role. Finally the coupling constants  reach  their  asymptotic  values  $\overline{\lambda}_*^{cl}$ and $\overline{\mu}_*^{cl}$ associated with  the classical  fixed point.  


\begin{figure}[!h]
\hskip  -0.6cm
\includegraphics[scale=0.53]{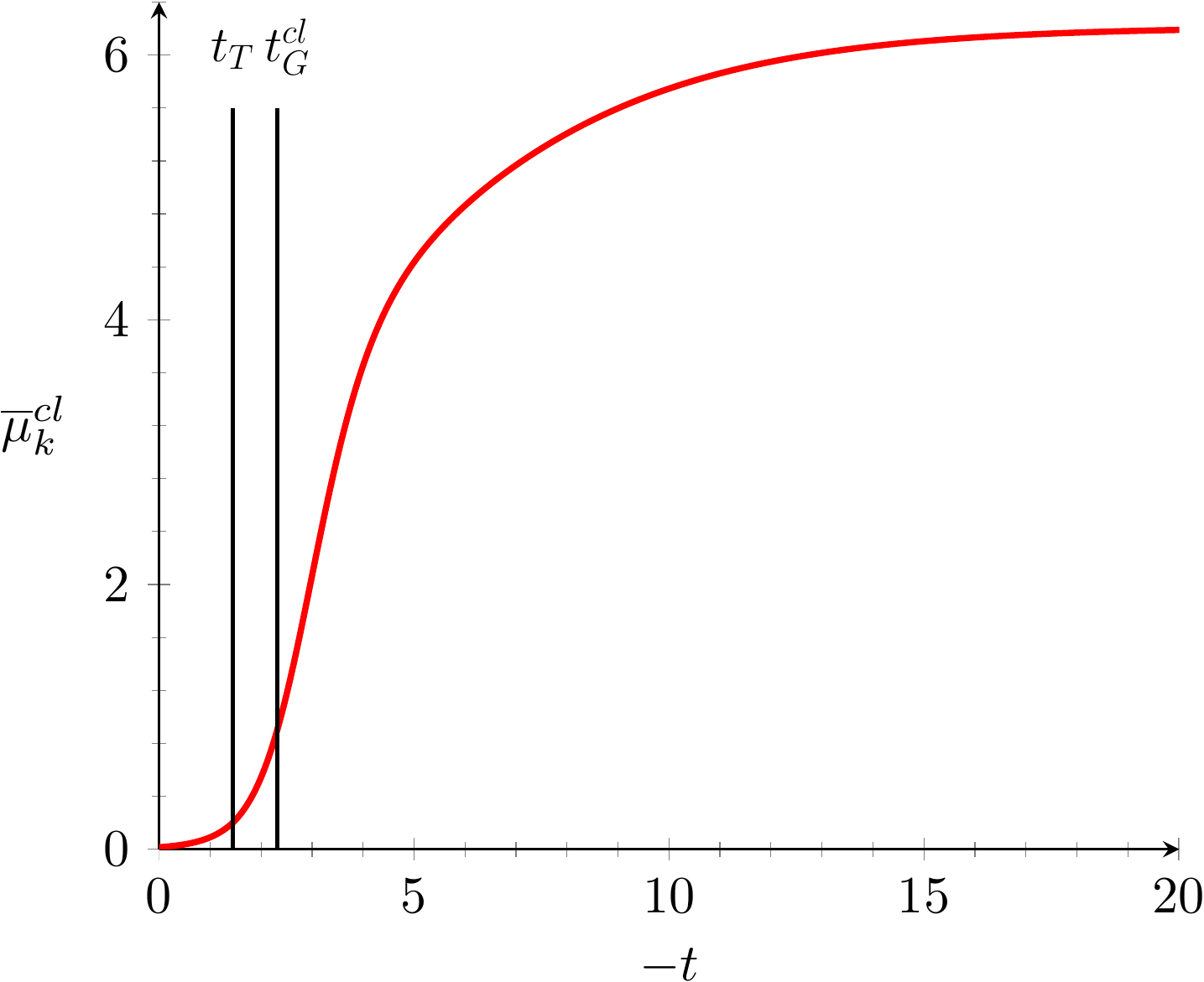}

\hskip  -0.8cm
 \includegraphics[scale=0.55]{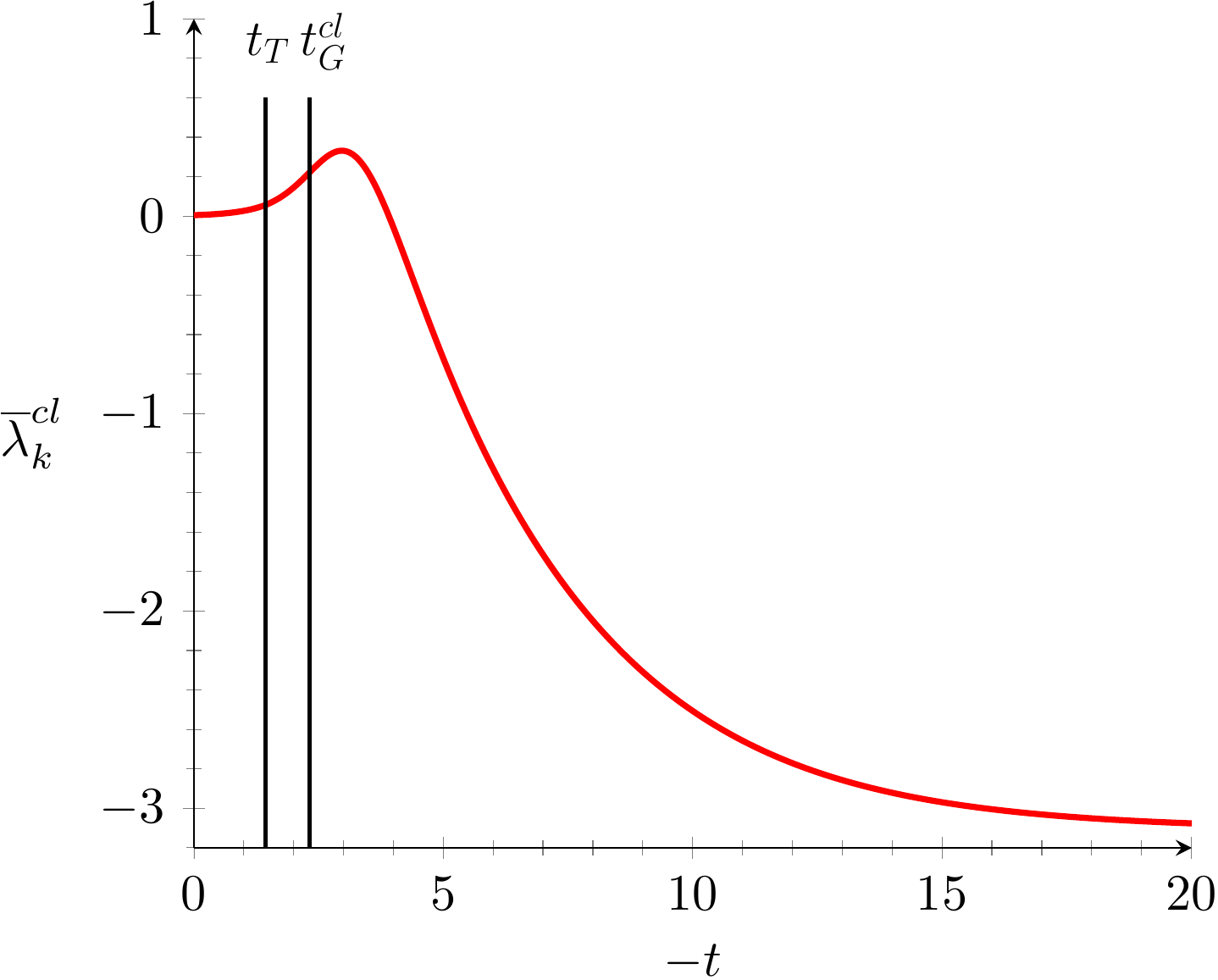}
\caption{Evolution of the dimensionless {\it classical}  coupling constants  $\overline\mu_k^{cl}$  and $\overline\lambda_k^{cl}$  with  RG time $-t=\ln \Lambda/k$.   $t_G^{cl}$ marks a clear crossover between a regime controlled by a classical gaussian fixed point
($\overline{\mu}_*^{cl}=\overline{\lambda}_*^{cl}=0$) and a regime controlled by the infra-red  non-trivial classical fixed point ($\overline{\mu}_*^{cl}\simeq6.21$, $\overline{\lambda}_*^{cl}\simeq-3.10$).}
\label{clcoup}
\end{figure}


\subsection{Crossover  behaviour of  physical quantities}

The previous   crossovers can also   be    observed for different relevant physical quantities:    the anomalous dimension $\eta_k$,  Fig.(\ref{EtaG}), the  dimensionless classical   bulk modulus $\overline K^{cl}_k=\overline\lambda^{cl}_k+\overline\mu^{cl}_k$, Fig.(\ref{bulk}), and the   dimensionless classical  Young  modulus $\overline{\mathcal{Y}}^{cl}_k=4 \overline\mu^{cl}_k(\overline\mu^{cl}_k+\overline\lambda^{cl}_k)/(2\overline\mu^{cl}_k+\overline\lambda^{cl}_k)$, Fig.(\ref{Young}).  They all share the same kind  evolution: relaxation toward vanishing values  in the quantum regime (not apparent on $K$ and $\mathcal{Y}$  due to the smallness of the    initial conditions), transition toward  a classical regime  and stabilization in the infra-red,  classical,  behaviour.  Note  the oscillations of $\eta_k$ at  early RG time $t$ on Fig.(\ref{EtaG})   due to the use of the discontinuous $\Theta$ cut-off. \\


\begin{figure}[!h]
\hskip -0.8cm
\includegraphics[scale=0.52]{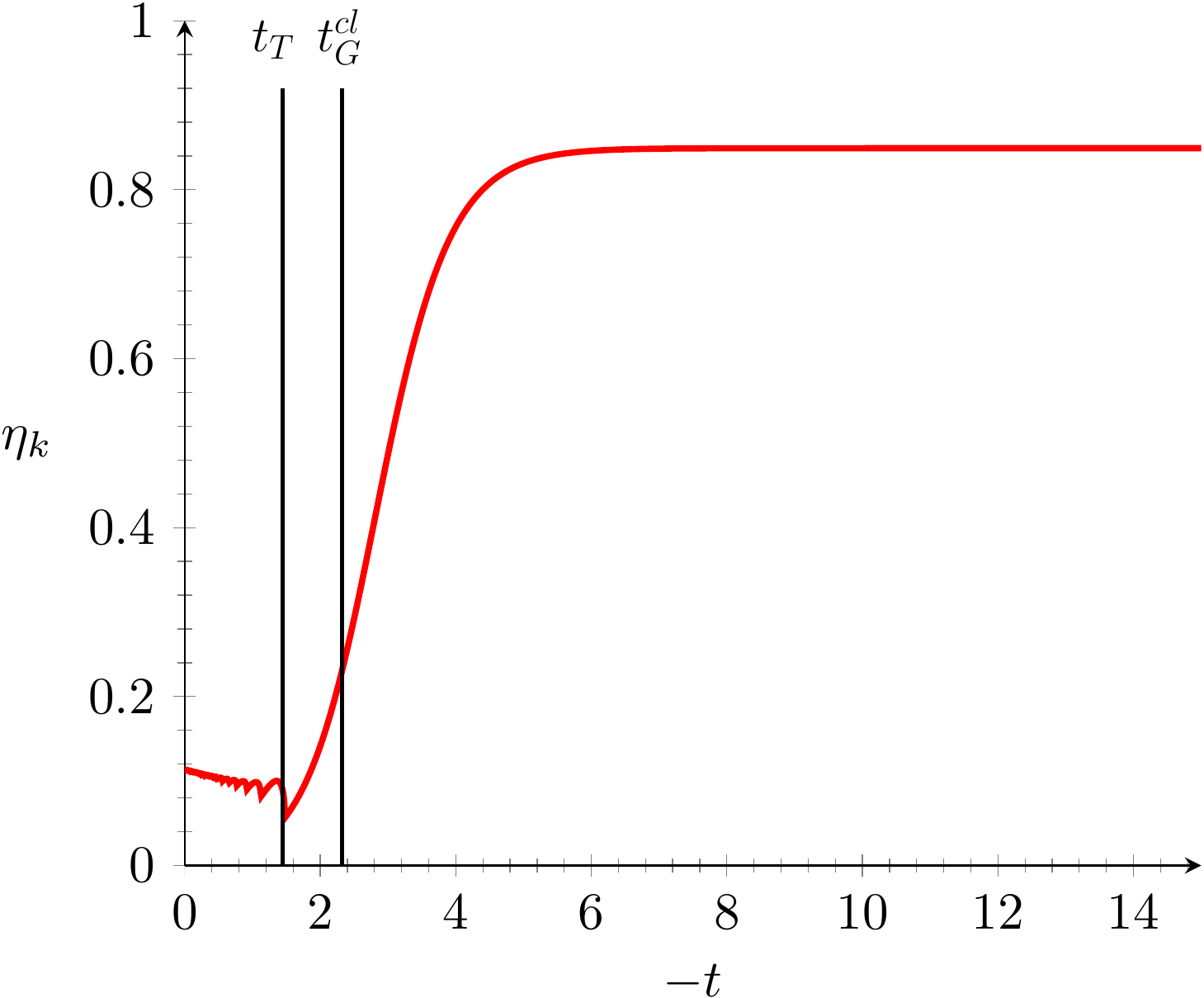}
\caption{ Evolution of the running anomalous dimension $\eta_k$ with  the RG time $-t=\ln \Lambda/k$.  $\eta_k$ evolves from  a quasi-gaussian (quantum) regime to a non-trivial   (classical)  regime where it  reaches  the fixed point  value $\eta^*=0.849$. The oscillations   at the beginning of the flow  are an artifact  related to the  use of  the $\Theta$ cut-off.}
 \label{EtaG}
\end{figure}

\begin{figure}[!h]
\hskip -1cm
\includegraphics[scale=0.5]{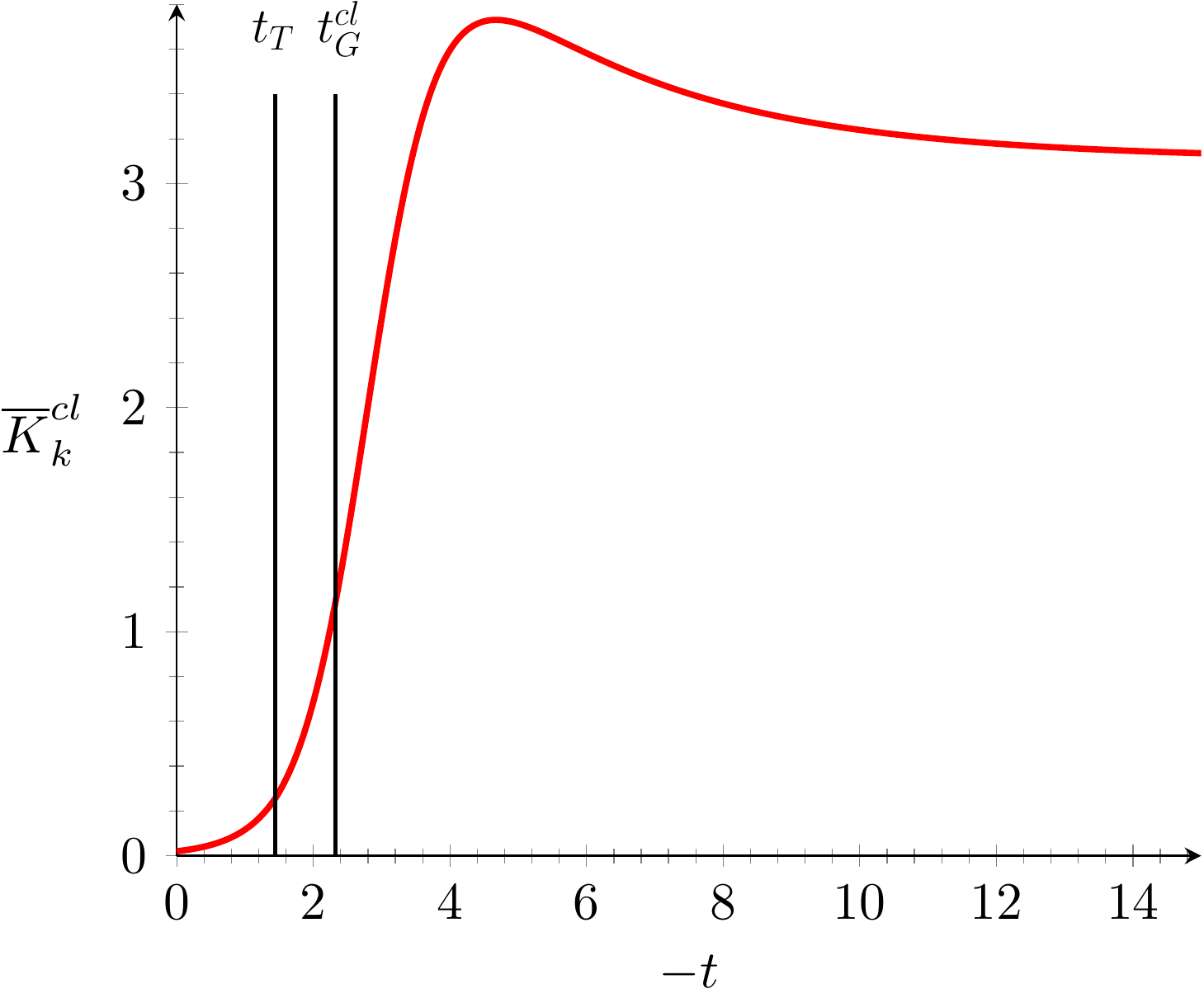}
\caption{  Evolution of  the dimensionless classical  bulk modulus $\overline K^{cl}_k$  the  with  the RG time $-t=\ln \Lambda/k$ that converges  to the value  $\overline K^{cl}_*\simeq 3.11$.   }.
\label{bulk}
\end{figure}


\begin{figure}[!h]
\hskip -1.5cm
\includegraphics[scale=0.5]{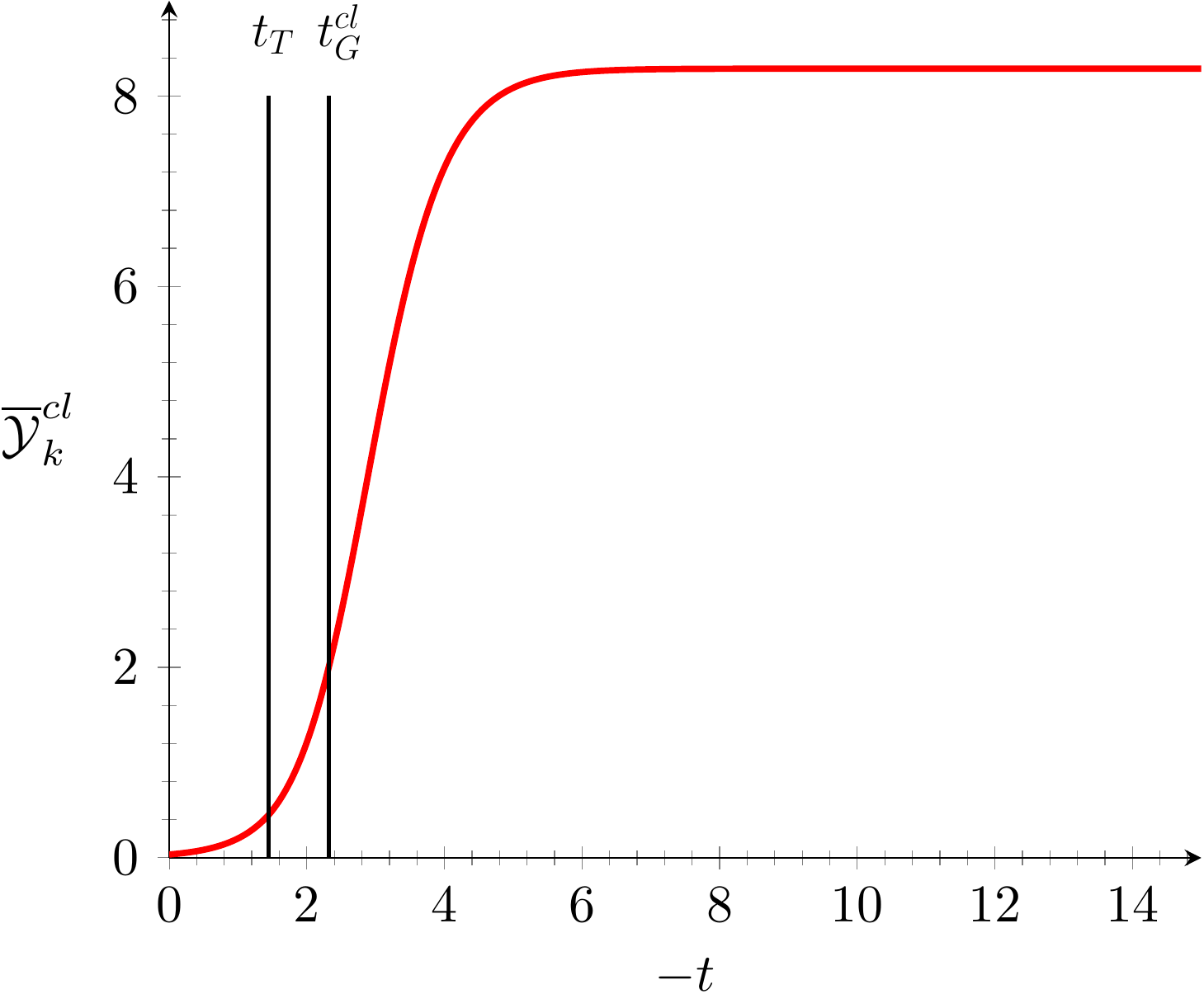}
\caption{Evolution of the dimensionless  classical Young modulus $\overline{\mathcal{Y}}^{cl}_k$  with the  RG time $-t=\ln \Lambda/k$  that converges  to the value $\overline{\mathcal{Y}}^{cl}_*\simeq 8.30$. }
\label{Young}
\end{figure}


Finally we have represented the Poisson ratio $\nu_k={\lambda_k}/({\lambda_k+2\mu_k})$, Fig.(\ref{PoissonG}).  Its evolution is  different from other quantities since, due to its expression  and the {\it "universal"} attractive line $\lambda=-\mu/2$,    it  does  converge   toward  the value  $\nu_k=-{1}/{3}$  in all regimes  of  temperatures. The main effect of the crossovers  is to change its  rate of convergence:  $\nu_k$ evolves  slowly in the quantum regime and more rapidly in the classical regime.


\begin{figure}[!h]
\hskip -1.5cm
\includegraphics[scale=0.55]{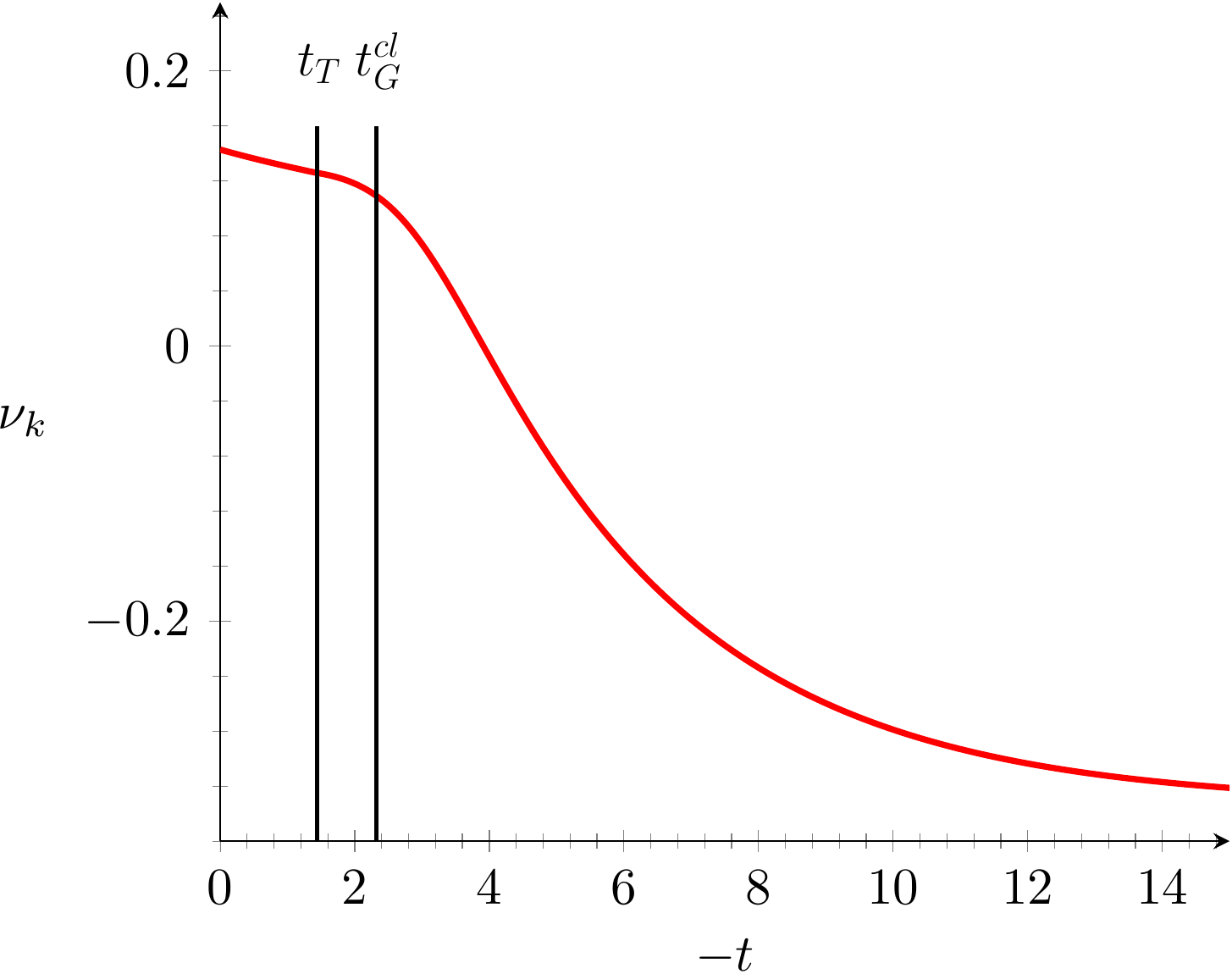}
\caption{Evolution of the Poisson ratio  $\nu_k$ with  the RG time $-t=\ln \Lambda/k$  that converges  to the value $\nu_*={-1/3}$.}
\label{PoissonG}
\end{figure}


\section{Conclusion}

 In this article we have  investigated  quantum polymerized membranes within  a NPRG   framework.  We have first derived, from  the  RG equations  describing {\it generic}  fluctuating  membranes, those  relevant to study the flat phase.  We have  then analyzed   the corresponding RG flow   that smoothly interpolates  between   a  quantum  and a classical regime. We have shown that, at vanishing temperatures,  the  infra-red behaviour is governed by an attractive  gaussian fixed point.    We have then  described the   quantum to classical crossover   leading to the usual non-trivial  infra-red  behaviour  associated with  stable classical membranes.      In each, quantum or classical, regime, we have exhibited the typical  momentum scale associated with  the crossover  between the weak and strong coupling regimes. Finally  we have illustrated these different  properties    in the case of free standing graphene. 

  We have  shown, in the context of the physics of membranes, the  ability  of the  NPRG approach  to continuously  interpolate between  a quantum regime governed by a gaussian fixed point and a classical regime governed by a strongly  interacting fixed point.  This relies on the nonperturbative character of  the  formalism used.  As emphasized  previously   approximations are nevertheless necessary and   their  nature   must be carefully adjusted  in accordance  with the physical context.    The present approach has been performed  at the  next-to-leading order   in the derivative expansion together with a field-expansion. Although  somewhat  preliminary  it has allowed us  to recover (and in some  cases  to contest)  previous  perturbative  --   weak-coupling or large $d$ --  results  and  to improve  them by  inclusion of  a nonvanishing anomalous  dimension and  nonperturbative  contributions.  
 
  Further and systematic improvements of  the present approach  could  be  implemented.   They   would consist, for instance,   in gradually  enriching  the effective action  $\Gamma_k$ used  by   increasing powers of   fields   and    field-derivatives.   A more qualitative  improvement would be  to consider  the whole  momentum dependence of the effective  action, an approach which has  been partially   performed in the classical case  \cite{braghin10,hasselmann11}  but  for which a  full treatment,  in  both the quantum and classical cases,  remains a challenge.    Such an   approach   would be particularly suitable to  the computation  of  thermodynamical quantities, like the thermal expansion coefficient or the specific heat  -- as   done  for instance in \cite{amorim14} -- since they  precisely  involve  the  full momentum content of  the two-point correlation  function.  It  would also  allow   to treat   the effects   of electron-phonon interaction   that  involves   a non-trivial  momentum   structure (see, e.g., \cite{guinea14}).

   From a methodological point of view  this  kind of approach  could   also allow to  clarify  the relation between the NPRG approach used  in this article and  the SCBA  approach performed  in   \cite{amorim14}  that lead to contradicting results.   We nevertheless  think that  this  disagreement  can   hardly   -- but not impossibly --  be  solved   by  considering  higher powers  of   field or field-derivatives as our result relies on the  rather strong evidence of the existence  of a  stable gaussian fixed point  at vanishing temperature.   In our opinion  this disagreement   suggests    to  reconsider  the arguments developed   within  the recent   controversy  involving  the authors of \cite{amorim14,amorim14b} and  those of \cite{kats14,kats14b} that precisely  concerns the way the anomalous dimension is computed in \cite{amorim14}.    More generally  and finally  an accurate control of   the  SCBA-SCSA   and  the NPRG approach   in the context of   quantum membranes    --  that appear,  in several  other contexts, notably that of disordered systems    (see for instance \cite{doussal03})   as  very   complementary   --  is highly    desirable.   \\

\acknowledgements

 We thank  B. Delamotte, P. Le Doussal,   N. Dupuis  and  J.-P. Kownacki   for helpful discussions and B. Amorim  who  provided us  very valuable  precisions about  the technique used in \cite{amorim14}.

\appendix

\section{Coupling constants definitions and  vertex functions}

\subsection{Coupling constants definitions} 
\label{coupling}

The coupling constants are defined as functional derivatives  of the effective action, considered in the configuration Eq.(\ref{rmin}): 
 \begin{widetext}
\begin{subequations}
\begin{align}
\mu_k&= \displaystyle   \lim_{\tilde{\bf p}\to 0}  \frac{1}{\zeta_k^2}\frac{d}{dp^2}\left[\Gamma^{(2)}_{k,D-1,D-1}[{\bf r}, {\bf \tilde p},-{\bf \tilde p}]\big |_{{\bf r}_{k,f}}\right] \label{defmu}\\
\nonumber \\
\lambda_k&=\displaystyle  \lim_{\tilde{\bf p}\to 0} \bigg\{\frac{1}{\zeta_k^2}\frac{d}{d(p_{D-1})^2} \left[\Gamma^{(2)}_{k,D-1,D-1}[{\bf r},{\bf \tilde p},-{\bf \tilde p}]\big |_{{\bf r}_{k,f}}\right]  - \frac{1}{\zeta_k^2}\frac{d}{dp^2}\left[\Gamma^{(2)}_{k,D-1,D-1}[{\bf r},{\bf \tilde p},-{\bf \tilde p}]\big |_{{\bf r}_{k,f}}  \right]  \bigg\}  \label{deflambda} \\
 \nonumber \\
Z_k&=\displaystyle   \lim_{\tilde{\bf p}\to 0} \frac{1}{2}\frac{d^2}{d(p^2)^2}\left[\Gamma^{(2)}_{k,D+1,D+1}[{\bf r},{\bf \tilde p},-{\bf \tilde p}]\big |_{{\bf r}_{k,f}}\right] \label{defz}  \\
\nonumber \\
Z^{\tau}_k&=\displaystyle \lim_{\tilde{\bf p}\to 0}   \frac{d}{d\omega_m^2}\left[\Gamma^{(2)}_{k,D+1,D+1}[{\bf r},{\bf \tilde p},-{\bf \tilde p}]\big |_{{\bf r}_{k,f}}\right]\ . 
\label{defztau}
\end{align}
\end{subequations}
\end{widetext}

The flow of $\zeta_k$  requires a specific treatment, see below. 

\subsection{RG flow of $\zeta_k$}

	Whereas the other coupling constants are defined from $\Gamma^{(2)}_k$, $\zeta_k$ is defined from the particular field configuration Eq.(\ref{rmin}).  One of its characterization is  that ${\bf r}_{k,f}$ is a minimum of the effective action:
\begin{equation}
\label{min}
\left.\frac{\delta\Gamma_k[\mathbf{r}]}{\delta\, r_j(\mathbf{\tilde q})}\right|_{\mathbf{r}_{k,f}}=0\ .
\end{equation}
However,  because of translational invariance of the action density, any flat configuration of the form:
\begin{equation}
r_{k,j}(\tilde {\bf q})=-i\sigma_k\,\delta_{n,0}\,\delta_{\gamma, j}\ \frac{\partial}{\partial q_{\gamma}}\delta({\bf q})
\end{equation}
   where  $\sigma_k$   represents  any real number not necessarily equal to $\zeta_k$  satisfies the condition Eq.(\ref{min}). This  last equation, therefore,   does not properly define   $\zeta_k$ as the extension parameter associated to the minimum of the effective action and cannot be used to compute its flow equation.  To  solve   this problem,  one has   to remove the ambiguity associated to  translational invariance. This can be done  by  replacing    the condition Eq.(\ref{min})  by  a condition where one performs  the   derivative  of   $\Gamma_k$ with respect to the tangent vectors  to the membrane:
\begin{equation}
\label{minn}
\left.\frac{\delta\Gamma_k[\mathbf{r}]}{\delta\,\partial_\alpha r_j(\mathbf{\tilde q})}\right|_{\mathbf{r}_{k,f}}=0 \ . 
\end{equation}

The flow of $\zeta_k$ is then deduced by taking a $t$-derivative of Eq.(\ref{minn}), and a little algebra.

\subsection{Vertex functions}
\label{Vertexfunctions}

The inverse propagator  in the  flat configuration  Eq.(\ref{rmin})  is  given by: 
\begin{equation}
\begin{array}{ll}
\Gamma^{(2)}_{k,i_1,i_2}[{\bf r}_{k,f};{\bf\tilde{q}}_1,{\bf\tilde{q}}_2]&= \delta({\bf\tilde{q}}_1+{\bf\tilde{q}}_2)\times \Big(G_F^{-1}(\tilde{\bf q}_1) {\cal P}_{i_1,i_2}^F(\tilde{\bf q}_1) \\
\\
&\hspace{-0.8cm}+G_{\perp}^{-1}(\tilde{\bf q}_1) {\cal P}_{i_1,i_2}^{\perp}(\tilde{\bf q}_1)+G_{\parallel}^{-1}(\tilde{\bf q}_1) {\cal P}_{i_1,i_2}^{\parallel}(\tilde{\bf q}_1)\Big)
\end{array}
\end{equation}
where we have defined  the flexural, transversal  and longitudinal projectors with  respect to momentum  ${\bf q}_1$ by:
\begin{subequations}
\begin{empheq}[left=\empheqlbrace]{align}
&{\cal P}_{i_1,i_2}^F(\tilde{\bf q}_1)=\delta_{i_1,i_2}-\delta_{\alpha, i_1}\delta_{\alpha,i_2}\\
\nonumber \\
&{\cal P}^{\perp}_{i_1,i_2}(\tilde{\bf q}_1)=\displaystyle \delta_{\alpha,i_1}\delta_{\alpha,i_2} -\delta_{\alpha,i_1}\delta_{\beta,i_2}\, {{q_{1,\alpha}}\, {q_{1,\beta}}\over { q}_1^2}\\
\nonumber  \\
&{\cal P}_{i_1,i_2}^{\parallel}(\tilde{\bf q}_1)=\displaystyle \delta_{\alpha,i_1}\delta_{\beta,i_2} {q_{1,\alpha}\,  q_{1,\beta}\over {q}_1^2}
\end{empheq}
\end{subequations}
and  where:
\begin{subequations}
\begin{empheq}[left=\empheqlbrace]{align}
&G_F^{-1}(\tilde{\bf q}_1)=Z_k\,{q}_1^4+Z^{\tau}_k\omega_{1n}^2\\
\nonumber\\
&G_{\parallel}^{-1}(\tilde{\bf q}_1)=Z_k\,{q}_1^4+Z^{\tau}_k\omega_{1n}^2+m_{2k}^2\,  q_1^2\\
\nonumber\\
&G_{\perp}^{-1}(\tilde{\bf q}_1)=Z_k\, {q}_1^4+Z^{\tau}_k\omega_{1n}^2+m_{1k}^2\,   q_1^2
\end{empheq}
\end{subequations}
with  $m_{1k}^2= \mu_k \zeta_k^2$ and  $m_{2k}^2=(2 \mu_k+\lambda_k)\zeta_k^2$.

As for the higher vertex functions  in the flat configuration Eq.(\ref{rmin})   there are given by: 
\begin{subequations}
\begin{align}
&\hspace{-0.2cm}\Gamma^{(3)}_{k,i_1,i_2,i_3}[{\bf r}_{k,f};{\bf\tilde{q}}_1,{\bf\tilde{q}}_2,{\bf\tilde{q}}_3]= i\, \zeta_k\, \delta({\bf\tilde{q}}_1+{\bf\tilde{q}}_2+{\bf\tilde{q}}_3)\times\nonumber \\
&\hspace{-0.2cm} \Big(\mu\big(({\bf q}_1.{\bf q}_2)\,  {\bf q}_{3,\alpha}+({\bf q}_1.{\bf q}_3)\,  {\bf q}_{2,\alpha}\big)\delta_{i_2,i_3}\delta_{\alpha,i_1} +{\hbox{perm.(1,2,3)}} \nonumber\\
&\hspace{-0.2cm} +\lambda\, ({\bf q}_1.{\bf q}_2)\,  {\bf q}_{3,\alpha}\, \delta_{i_1,i_2}\delta_{\alpha,i_3}+{\hbox{perm.(1,2,3)}}\Big) \label{gamma3}
\\
\nonumber \\
&\hspace{-0.3cm}\Gamma^{(4)}_{k,i_1,i_2,i_3,i_4}[{\bf r}_{k,f};{\bf\tilde{q}}_1,{\bf\tilde{q}}_2,{\bf\tilde{q}}_3,{\bf\tilde{q}}_4]=\delta({\bf\tilde{q}}_1+{\bf\tilde{q}}_2+{\bf\tilde{q}}_3+{\bf\tilde{q}}_4)\times \nonumber  \\
&\hspace{-0.3cm} \Big(\mu\, ({\bf q}_1.{\bf q}_2)\,  ({\bf q}_3.{\bf q}_4)  (\delta_{i_1,i_4}\delta_{i_2,i_3}+\delta_{i_1,i_3}\delta_{i_2,i_4})+{\hbox{perm.(1,2,3,4)}}\nonumber  \\
&\hspace{-0.3cm} +\lambda({\bf q}_1.{\bf q}_2)\,  ({\bf q}_3.{\bf q}_4)\, \delta_{i_1,i_2}\delta_{i_3,i_4}+ {\hbox{perm.(1,2,3,4)}}\Big)\label{gamma4}\ .
\end{align}
\end{subequations}

\section{Thresholds functions}
\label{threshold}

The  RG flow equations Eqs.(\ref{eqrg1})-(\ref{eqrg3}),(\ref{flot1})-(\ref{flot2})  and (\ref{floteta})    are expressed in terms of two families of  threshold functions $l_{abc}^D$ and $n_{abc}^D$ defined   in Eq.(\ref{thresholdL}) and (\ref{thresholdN}) that contain all the remnants of the loop integration of Eq.(\ref{renorm}).  In this Appendix, we write their dimensionless counterparts and  compute them explicitly for the $\Theta$ cut-off.

 \subsection{Dimensionless threshold functions}
 \label{Dimthresfunct}

We express the integrals Eq.(\ref{thresholdL}) and Eq.(\ref{thresholdN})  in terms of the dimensionless variables $y=q^2/k^2$ and $\overline{\omega}_n=\omega_n/(\Delta_k k^2)$. We also  introduce the variable: $Y=y^2+\overline{\omega}_n^2$ so that the cut-off  function $R_k(\tilde{\bf  q})$  { entering in the expression (\ref{cutoff2}) reads:
\begin{equation}
R_k(\tilde{\bf  q})=Z_k\, k^4\, Y\, {\cal R}(Y)\ .
\label{GenCut}
\end{equation}
\\
 The  two families of threshold functions  $\overline{l}^D_{abc}$ and $\overline{n}^D_{abc}$   involve  $\partial_tR_k(\tilde{\bf  q})$.   This last quantity  can be made  explicit  thanks to the following relations (that take account  of the fact that  $\tilde\eta_k=0$):
\begin{equation}
\begin{array}{ll}
\partial_t \Delta_k&=-\displaystyle{1\over 2}\eta_k\Delta_k\\
\\
\partial_t  Y& =-4Y+\overline{\omega}_n^2\eta_k\ .
\end{array} 
\end{equation}
This leads to, with  the notations ${\cal R}={\cal R}(Y)$, ${\cal R}'={\cal R}'(Y)$:
\begin{equation}
\partial_t R_k(\tilde{\bf  q})=-Z_k\, k^4\Big(4Y^2\,{\cal R}'+\eta_k(y^2{\cal R}-\overline{\omega}_n^2Y\,{\cal R}')\Big)\ . 
\label{dtR}
\end{equation}
As seen in the expression (\ref{thresholdN})  the  threshold  function family $n_{abc}^D$ requires to compute additionally   the quantity 
$\widehat\partial_t \partial_{q^2}P(\tilde{\bf q})$  where we recall that $P(\tilde{\bf q})=Z_k\, q^{4}+Z_k^{\tau} \omega_n^2+ R_{k}(\tilde{\bf q})$.  Since the operator $\widehat\partial_t$ only acts on $R_k(\tilde{\bf  q})$ one has the identity:  $\widehat\partial_t \partial_{q^2}P(\tilde{\bf q})=\partial_{q^2}\partial_t  R(\tilde{\bf q})$.  This last quantity is easily computed from the expression Eq.(\ref{dtR}) and one gets:  
\begin{align}
\widehat\partial_t \partial_{q^2}P(\tilde{\bf q})&=-2Z_k k^2y\Big(4Y(2{\cal R}'+Y{\cal R}'')   \nonumber \\
& +\eta_k\big({\cal R}+Y{\cal R}'- 2\overline{\omega}^2_n{\cal R}'-Y\overline{\omega}^2_n {\cal R}''\big)\Big)\  . 
\end{align} 

Finally, we get the expressions of the dimensionless threshold functions  $\overline{l}_{abc}^D$ and $\overline{n}_{abc}^D$  defined by: 	\begin{subequations}
\begin{empheq}[left=\empheqlbrace]{align}
l_{abc}^D&=(Z_kk^4)^{1-a-b-c}\,k^{D+2}\Delta_k\:\overline{l}_{abc}^D\\
\nonumber\\
n_{abc}^D&=(Z_kk^4)^{1-a-b-c}\,k^{D+2}\Delta_k\:\overline{n}_{abc}^D
\end{empheq}
\end{subequations}

\begin{widetext}
\begin{subequations}
\begin{align}
\overline{l}^D_{abc}&=\displaystyle -\frac{\overline{T}_k}{2}\sum_{\overline{\omega}_n}\int_0^{\infty} dy\:y^{\frac{D}{2}-1}\,
\frac{4Y^2\,{\cal R}'+\eta_k(y^2{\cal R}-\overline{\omega}_n^2Y\,{\cal R}')}{{[\overline P_0(Y)]}^{a} {[\overline P_1(Y)]}^{b} {[\overline P_2(Y)]}^{c}}
\Big\{\frac{a}{\overline P_0(Y)}+\frac{b}{\overline P_1(Y)}+\frac{c}{\overline P_2(Y)}\Big\} \label{ldedim0}  \\
\nonumber \\
\overline{n}^D_{abc}&=\displaystyle \overline{T}_k\sum_{\overline{\omega}_n}\int_0^{\infty}dy \frac{y^{\frac{D+2}{2}}}{{[\overline P_0(Y)]}^{a} {[\overline P_1(Y)]}^{b}
{[\overline P_2(Y)]}^{c}}\bigg\{\Big(4Y(2{\cal R}'+Y{\cal R}'')+\eta_k({\cal R}+Y{\cal R}'-2\overline{\omega}^2_n{\cal R}'-Y\overline{\omega}^2_n{\cal R}'')\Big) \nonumber \\
& \hspace{0.5cm} \displaystyle  + \big(1+{\cal R}+Y{\cal R}'\big)\big(4Y^2\,{\cal R}'+\eta_k(y^2{\cal R}-\overline{\omega}_n^2Y\,{\cal R}')\big)\Big\{\frac{a}{\overline P_0(Y)}+\frac{b}{\overline P_1(Y)}+\frac{c}{\overline P_2(Y)} \Big\} \bigg\} \label{ndedim0}
\end{align}
\end{subequations}
\end{widetext}
with $\overline{P}_i(Y)=\overline{P}(Y)+\overline{m}_{ik}^2\, y$, $i=0,1,2$.

\subsection{Threshold functions with $\Theta$ cut-off}
\label{thresholdtheta}

The $\Theta$ cut-off  is particularly useful if one  wants  to get simple analytical results.  Using the expression Eq.(\ref{cutoff2}), one has:
\begin{equation}
{\cal R}(Y)=\frac{1-Y}{Y}\Theta(1-Y)
\label{thetacut}
\end{equation}
where $\Theta$ is the Heaviside step function. The  derivatives of ${\cal R}$  with respect to $Y$ are  needed:
\begin{subequations}
\begin{align}
&{\cal R}'(Y)=-\displaystyle\frac{1}{Y^2}\Theta(1-Y)-\frac{1-Y}{Y}\delta(1-Y) \\
\nonumber \\
& {\cal R}''(Y)=\displaystyle \frac{2}{Y^3}\Theta(1-Y)+\frac{2}{Y^2}\delta(1-Y)-\frac{1-Y}{Y}\delta'(1-Y) \ . 
\end{align}
\end{subequations}

Thanks to the properties of $\Theta$ and $\delta$ functions  several terms  simplify  in the expressions  of the threshold functions Eqs.(\ref{ldedim0})  and (\ref{ndedim0}).
 Also, the $\Theta$ function makes  all remaining sums and integrals finite.  We recall that the sums are cut at a maximal frequency given by  $n_{max}=\lfloor1/(2\pi\overline{T}_k)\rfloor$:
\begin{subequations}
\begin{align}
&\overline{l}^D_{abc}=\frac{\overline{T}_k}{2}\hspace{-0.2cm}\sum_{n=-n_{max}}^{n_{max}}\hspace{-0.2cm} \int_0^{\sqrt{1-\overline{\omega}_n^2}} dy\:y^{\frac{D}{2}-1}\, \frac{4+\eta_k(y^2-1)}{{[1+\overline{m}_{1k}^2y]}^{b} {[1+\overline{m}_{2k}^2y]}^{c}} \nonumber\\
& \hspace{2.5cm} \times\bigg\{a+\frac{b}{1+\overline{m}_{1k}^2y}+\frac{c}{1+\overline{m}_{2k}^2y}\bigg\} \label{ldedim} \\
 \nonumber\\
&\overline{n}^D_{abc}=-\overline{T}_k \sum_{n=-n_{max}}^{n_{max}}\Bigg\{\int_0^{\sqrt{1-\overline{\omega}_n^2}}dy\:\frac{\eta_k\ y^{\frac{D+2}{2}}}{{[1+\overline{m}_{1k}^2y]}^{b}{[1+\overline{m}_{2k}^2y]}^{c}} \nonumber\\
&\hspace{1cm}  -\frac{(4-\eta_k\overline{\omega}_n^2)(1-\overline{\omega}_n^2)^{\frac{D}{4}}}{2{\Big[1+\overline{m}_{1k}^2\sqrt{1-\overline{\omega}_n^2}\Big]}^b{\Big[1+\overline{m}_{2k}^2\sqrt{1-\overline{\omega}_n^2}\Big]}^c}\Bigg\} \label{ndedim}\  .
\end{align}
\end{subequations}

\subsection{Flat phase threshold functions}

In the flat phase, only massless threshold functions $\overline{l}^D_{a00}$ and $\overline{n}^D_{a00}$ remain   for which 
 the   $y$-integrals in Eq.(\ref{ldedim}) and (\ref{ndedim}) can be computed explicitly and yield:
\begin{subequations}
\begin{align}
&\overline{l}^D_{a00}=\overline{T}_k\hspace{-0.3cm}\sum_{n=-n_{max}}^{n_{max}}\hspace{-0.2cm}\bigg[4\frac{a}{D}(1-\overline{\omega}_n^2)^{\frac{D}{4}}-\eta_k\frac{a(4+D\,\overline{\omega}_n^2)}{D(D+4)}(1-\overline{\omega}_n^2)^{\frac{D}{4}}\bigg] \label{flatl}
\\
&\overline{n}^D_{a00}=\overline{T}_k\hspace{-0.2cm}\sum_{n=-n_{max}}^{n_{max}}\hspace{-0.1cm}\bigg[2(1-\overline{\omega}_n^2)^{\frac{D}{4}}-\eta_k\frac{4+D\,\overline{\omega}_n^2}{2(D+4)}(1-\overline{\omega}_n^2)^{\frac{D}{4}}\bigg] \ . \label{flatn}\ 
\end{align}
\end{subequations}

 As long as $2\pi\overline{T}_k<1$, $n_{max}>0$} and the frequency sum brings several contributions.

\subsubsection{Vanishing temperature limit}
			
When $\overline{T}_k\rightarrow0$, one recognises a Riemann integral:
\[\overline{T}_k\sum_{n=-n_{max}}^{n_{max}}\:\underset{\overline{T}_k\rightarrow0}{\longrightarrow}\:\int_{-1}^1\frac{d\overline{\omega}}{2\pi}\ .\]

The threshold functions now reads:
\begin{subequations}
\begin{empheq}[left=\empheqlbrace]{align}
\overline{l}^D_{a00}=\frac{a}{2\sqrt{\pi}}\frac{\Gamma\left[\frac{D}{4}\right]}{\Gamma\left[\frac{6+D}{4}\right]}-\eta_k\frac{3a}{16\sqrt{\pi}}\frac{\Gamma\left[\frac{D}{4}\right]}{\Gamma\left[\frac{10+D}{4}\right]} \label{ldplate0} \\
\nonumber\\
\overline{n}^D_{a00}=\frac{1}{\sqrt{\pi}}\frac{\Gamma\left[\frac{4+D}{4}\right]}{\Gamma\left[\frac{6+D}{4}\right]}-\eta_k\frac{3}{8\sqrt{\pi}}\frac{\Gamma\left[\frac{4+D}{4}\right]}{\Gamma\left[\frac{10+D}{4}\right]} \label{ndplate0}\ .
\end{empheq}
\end{subequations}

These expressions are valid only at $\overline{T}_k=0$ strictly speaking.
 However   they  still provide a very good approximation to the value of the threshold functions in the whole region $\overline{T}_k\ll1$.

 \subsubsection{High temperature limit}
 
At high-temperatures,  the only remaining term in the frequency sum corresponds  to $\overline{\omega}_0=0$. The threshold functions become:
\begin{subequations}
\begin{empheq}[left=\empheqlbrace]{align}
&\overline{l}^D_{a00}=\displaystyle \overline{T}_k\bigg[\frac{4a}{D}-\eta_k\frac{4a}{D(D+4)}\bigg] \label{flathigha} \\
\nonumber \\
&\overline{n}^D_{a00}=\displaystyle  \overline{T}_k\bigg[2-\eta_k\frac{2}{D+4}\bigg]\ .\label{flathighb}
\end{empheq}
\end{subequations}

The $\overline{T}_k$ factor in the flow equations is combined with the coupling constants to give back their value in the classical theory. 

\vspace{1cm}


\end{document}